\documentclass[preprint,sort&compress,12pt]{elsarticle}
\usepackage{bm} % required for: \bm{}
\usepackage{amsfonts} % required for: \mathbb{}, \mathcal{}, ...

\newcommand{\Fcal}{\ensuremath{\mathcal{F}}}
\newcommand{\Gcal}{\ensuremath{\mathcal{G}}}

\newcommand{\Ncal}{\ensuremath{\mathcal{N}}}

\newcommand{\Pcal}{\ensuremath{\mathcal{P}}}

\newcommand{\Scal}{\ensuremath{\mathcal{S}}}

\newcommand{\Xboldcal}{\ensuremath{\boldsymbol{\mathcal{X}}}}

\newcommand\Ibm{{\ensuremath{\bm{I}}}}

\newcommand\Sbm{{\ensuremath{\bm{S}}}}

\newcommand\Wbm{{\ensuremath{\bm{W}}}}

\newcommand\Ybm{{\ensuremath{\bm{Y}}}}

\newcommand\nbm{{\ensuremath{\bm{n}}}}

\newcommand\ubm{{\ensuremath{\bm{u}}}}

\newcommand\xbm{{\ensuremath{\bm{x}}}}

\newcommand\ubold{\ensuremath{\mathbf{u}}}

\newcommand\mubold{{\ensuremath{\boldsymbol{\mu}}}}

\newcommand\xibold{{\ensuremath{\boldsymbol{\xi}}}}

\newcommand\epsilonbold{{\ensuremath{\boldsymbol{\epsilon}}}}

\newcommand\Psibold{{\ensuremath{\boldsymbol{\Psi}}}}

\usepackage{tikz}
\usepackage{pgfplots}
\usepackage{pgfplotstable, filecontents, booktabs}
\pgfplotsset{compat=1.9}

\usetikzlibrary{pgfplots.groupplots}
\usepgfplotslibrary{fillbetween}
\usetikzlibrary{calc,fit,matrix,arrows,automata,positioning,shapes}
\usetikzlibrary{arrows.meta}

\pgfplotsset{select coords between index/.style 2 args={
    x filter/.code={
        \ifnum\coordindex<#1\fi
        \ifnum\coordindex>#2\fi
    }
}}

\tikzset{
 invisible/.style={opacity=0},
 visible on/.style={alt={#1{}{invisible}}},
 alt/.code args={<#1>#2#3}{%
   \alt<#1>{\pgfkeysalso{#2}}{\pgfkeysalso{#3}}
 },
}

\usepackage{amssymb}
\usepackage{amsthm}
\usepackage{amsmath}
\usepackage{mathrsfs}
\usepackage{mathtools}
\usepackage{graphicx}
\usepackage[font=normalsize,labelfont=bf]{caption}
\usepackage[font=normalsize]{subfig}
\usepackage{algorithm}
\usepackage{algpseudocode}
\usepackage{array}
\usepackage{multirow}
\usepackage{enumerate}
\usepackage{diagbox} 
\theoremstyle{remark}

\usepackage{tikz}
\usepackage[utf8]{inputenc}
\usepackage{pgfplots}
\pgfplotsset{compat=newest}
\usepgfplotslibrary{groupplots}
\usepgfplotslibrary{dateplot}
\usepackage{tikzscale}
\usetikzlibrary{arrows,automata}
\setcitestyle{square}
\usepackage{lineno}
\usepackage{fullpage}
\usepackage{xcolor}
\usepackage[colorinlistoftodos]{todonotes}
\usepackage{bbm}
\usepackage{listings}
\usepackage[colorlinks=true]{hyperref}
\usepackage{url}

\graphicspath{ {./figs/} }
\biboptions{comma,round}

\graphicspath{ {./figs/} }

\journal{Elsevier}

\DeclareUnicodeCharacter{2212}{-} 
\begin{document}
\begin{frontmatter}
%\tableofcontents
% \listoftodos

%% Title, authors and addresses
 \title{Super-resolution and denoising of fluid flow using physics-informed convolutional neural networks without high-resolution labels}

%% use the tnoteref command within \title for footnotes;
%% use the tnotetext command for the associated footnote;
%% use the fnref command within \author or \address for footnotes;
%% use the fntext command for the associated footnote;
%% use the corref command within \author for corresponding author footnotes;
%% use the cortext command for the associated footnote;
%% use the ead command for the email address,
%% and the form \ead[url] for the home page:
%%
%% \title{Title\tnoteref{label1}}
%% \tnotetext[label1]{}
%% \author{Name\corref{cor1}\fnref{label2}}
%% \ead{email address}
%% \ead[url]{home page}
%% \fntext[label2]{}
%% \cortext[cor1]{}
%% \address{Address\fnref{label3}}
%% \fntext[label3]{}

%% use optional labels to link authors explicitly to addresses:
%% \author[label1,label2]{<author name>}
%% \address[label1]{<address>}
%% \address[label2]{<address>}

\author[ndAME]{Han Gao}
\author[ndAME]{Luning Sun}
\author[ndAME]{Jian-Xun Wang\corref{corxh}}

\address[ndAME]{Department of Aerospace and Mechanical Engineering, University of Notre Dame, Notre Dame, IN}
%\address[ndCICS]{Center for Informatics and Computational Science, University of Notre Dame, Notre Dame, IN}
%
\cortext[corxh]{Corresponding author. Tel: +1 540 3156512}
\ead{jwang33@nd.edu}

\begin{abstract}
%%135 words, 964 characters, 150 words are the maximum for some journals!%%
High-resolution (HR) information of fluid flows, although preferable, is usually less accessible due to limited computational or experimental resources. In many cases, fluid data are generally sparse, incomplete, and possibly noisy. How to enhance spatial resolution and decrease the noise level of flow data is essential and practically useful. Deep learning (DL) techniques have been demonstrated to be effective for super-resolution (SR) tasks, which, however, primarily rely on sufficient HR labels for training. In this work, we present a novel physics-informed DL-based SR solution using convolutional neural networks (CNN), which is able to produce HR flow fields from low-resolution (LR) inputs in high-dimensional parameter space. By leveraging the conservation laws and boundary conditions of fluid flows, the CNN-SR model is trained without any HR labels. Moreover, the proposed CNN-SR solution unifies the forward SR and inverse data assimilation for the scenarios where the physics is partially known, e.g., unknown boundary conditions. Several flow SR problems relevant to cardiovascular applications have been studied to demonstrate the proposed method's effectiveness and merit. Both Gaussian and non-Gaussian MRI noises are investigated to illustrate the denoising capability.

%High-resolution spatial descriptions of flow are essential for understanding fluid phenomena. However, it can be challenging to obtain such descriptions experimentally or numerically in terms of data sparsity and high computational cost. We present a novel deep learning framework that can generate deterministic/parametric high-resolution fields from low-resolution ones. By leveraging the physics knowledge, the convolutional neural network (CNN) is trained without high-resolution labeled data. Moreover, a new way to assimilate sparse observations to super-resolve flow with unknown boundary conditions is proposed. To deal with potential noisy input, we take into account noise in the offline phase that can be generalized to the online phase. We perform several experiments, including one in a Magnetic Resonance Imaging (MRI) related scenario and one in a high-dimensional situation to demonstrate the efficiency, accuracy, and flexibility of the framework.  
\end{abstract}

\end{frontmatter}

%\linenumbers
% \input{highlights}
 \clearpage
\section{Introduction}
\label{sec:intro}
High-resolution (HR) information of fluid flow is critical for reliable qualitative and quantitative analyses for fluid systems in aerodynamics, mechanical, and biomedical engineering. Nonetheless, fluid flow data are often \emph{sparse, incomplete, and noisy} in real-life scenarios due to the following reasons. \underline{First}, flow data are typically spatiotemporal fields in large scales, which poses significant challenges to data analysis, sharing, and visualization due to limited storage space and large communication overhead. For example, the direct numerical simulation (DNS) of wall-bounded turbulent flows at Reynolds number of $Re_{\tau} = 10^4$ can generate more than 20 TB files at each time step, and the file size will increase exponentially as $Re_{\tau}$ grows~\cite{pollard2016whither}. Hence, scientists could only afford to store a small fraction of data (e.g., temporally sparse sequences, spatially downscaled volumes, or selective variable subsequences) for post hoc analysis. \underline{Second}, the data resolution is often constrained by the ability of the measurement techniques. For example, flow magnetic resonance (MR) imaging has been widely used to quantitatively study cardiovascular blood flow dynamics~\cite{lawley20184d,stankovic20144d}, but the spatial resolution and signal-to-noise ratio (SRN) of flow MR data are far from sufficient, limiting their clinical applications~\cite{ong2015robust,Fathi:2018fv,Callaghan:2017jm}. Therefore, it is significant and imperative to enhance the resolution and reduce the noise level, which is referred to as flow data \emph{super-resolution and denoising}. \underline{Moreover}, fluid flow data, even with high spatiotemporal resolutions, are usually sparse in the parameter space due to limited computational or experimental resources. For example, a single run of fully-resolved DNS of turbulent flows often takes days or weeks on high-performance computing facilities~\cite{pollard2016whither}. It becomes infeasible to perform massive queries in the parameter space to explore many different boundaries, geometries, and operational configurations for uncertainty quantification (UQ) and optimization. In such scenarios, data super-resolution in the parameter space can be treated as a cost-effective \emph{surrogate model} that leverages the use of efficient low-resolution (LR) simulations or experiments. Scientists can opt to run their simulations or experiments at a low resolution and then upscale the results back to the target resolution, which will significantly save cost and speed up the process of scientific investigation and discovery. 

Various efforts have been devoted to enhancing the spatial or/and temporal resolution of fluid flows. One type of approach focuses on extracting the coherent structures and correlation features from an existing HR database based on proper orthogonal decomposition (POD)~\cite{venturi2004gappy,bui2003proper,podvin2006reconstruction,yakhot2007reconstruction,moreno2016aerodynamic,mifsud2019fusing}, dynamic mode decomposition (DMD)~\cite{schmid2010dynamic,tu2013dynamic}, or other sparsity-promoting representation techniques~\cite{callaham2019robust,manohar2018data}. However, these approaches are limited by the linearity assumption made for the reduced basis. The other types of super-resolution methods take advantage of the computational fluid dynamics (CFD) model to provide full-field predictions instead of learning from the offline database. The sparse LR data are fused into the CFD predictions using data assimilation (DA) techniques, e.g., ensemble Kalman filter, particle filters, or variational DA algorithms~\cite{foures2014data,combes2015particle,kikuchi2015assessment,mons2016reconstruction,symon2017data,wang2016data,xiao2016quantifying}. Nonetheless, physics-based CFD simulations are time-consuming in general, while the DA process usually involves numerous model evaluations, which could be computationally prohibitive.

The recent advances in machine learning (ML) and GPU computing open up a promising revenue to tackle this challenge. In past a few years, ML has been successfully applied in fluid dynamics~\cite{brunton2020machine,brenner2019perspective}, for, e.g., turbulence closure modeling~\cite{ling2016reynolds,duraisamy2019turbulence,wang2017physics,wang2019prediction,zafar2020convolutional}, inflow turbulence generation~\cite{fukami2019synthetic,kim2020deep}, and fluid surrogate/reduced-order modeling~\cite{sun2020surrogate,maulik2020probabilistic,gao2019non}, etc. In particular, the growing success of deep learning (DL)-based image super-resolution~\cite{ledig2017photo} in computer vision inspires the application of deep neural networks (DNN) for the flow field super-resolution and reconstruction~\cite{fukami2020machine,fukami2019super, fukami2018super, liu2020multiresolution, deng2019super, bode2019using, bai2019dynamic, gonzalez2018deep, xie2018tempogan}. Fukami et al.~\cite{fukami2020machine,fukami2019super,fukami2018super} applied the convolutional neural network (CNN) and hybrid downsampled skip-connection multiscale (DSC/MC) models for super-resolving downsampled HR data of both laminar and turbulent flows. To achieve a similar goal, Deng et al.~\cite{deng2019super} applied generative adversarial networks (GAN), while Liu et al.~\cite{liu2020deep} adopted multiple temporal paths convolutional neural network (MTPC). Thuerey and co-workers designed a more complicated GAN architecture by considering temporal coherence to up-sample 3-D volumetric turbulent smoke data~\cite{xie2018tempogan}, and they further improved the scalability by decomposing the learning problem into multiple smaller sub-problems~\cite{werhahn2019multi}. Bai et al~\cite{bai2019dynamic} used a dictionary learning strategy to super-resolve turbulent smoke flows in a variety of animation context. Guo et al.~\cite{guo2020ssr} designed DL-based spatial upscaling solutions of vector fields for visualization purposes. Considering multi-scale features in fluid dynamics, Liu et al.~\cite{liu2020multiresolution} proposed a multi-resolution convolutional autoencoder (MrCAE) super-resolution architecture to dynamically capture different scaled flow features at different depths of the network, where the multi-grid method and transfer learning techniques are leveraged. Instead of using deep networks, Erichson et al.~\cite{erichson2019shallow} proposed to directly capture an end-to-end mapping between the sparse measurements and the HR flow field using a shallow network. In the context of biomedical imaging, Ferdian et al.~\cite{ferdian20204dflownet} developed a DL model for 4D flow MRI super-resolution, where CFD simulation data are utilized as HR labels for training.

Despite the great promise, the success of these DL models mainly relies on a large amount of offline HR data as labels, which are inaccessible in many cases, e.g., super-resolution of 4D flow MR images. Moreover, these recent data-driven upsampling approaches add visual complexity to an LR input but cannot guarantee that the super-resolved fields are faithful to the physical laws and principles. A more promising strategy is to incorporate prior physics knowledge into deep learning models to alleviate data requirements and improve learning performance. This idea of physics-informed deep learning has been recently explored for solving forward and inverse PDEs~\cite{raissi2019physics,dwivedi420distributed}, surrogate modeling~\cite{sun2020surrogate,zhu2019physics,geneva_modeling_2020,zhang2020physics}, and equation discovery~\cite{long2017pde,long2019pde,singh2020time,chen2020deep}. For flow reconstruction and super-resolution, the divergence-free constraint for incompressible flow is the most straightforward one to be imposed on the learned solution, which can be done in a hard manner by either introducing stream functions or using spectral methods~\cite{jiang2020meshfreeflownet,mohan2020embedding,bode2019using,thuereyphysics}. Jiang et al.~\cite{jiang2020meshfreeflownet} proposed a MeshfreeFlowNet for super-resolving of LR solution fields of Rayleigh–Benard convection equations, where the training is regularized by the governing PDEs. In a similar vein, Subramaniam et al.~\cite{subramaniam2020turbulence} utilized the mass and momentum conservation law to constrain the training of a GAN for turbulence enrichment. Sun and Wang~\cite{sun2020physics} developed a Bayesian physics-informed neural network using Navier-Stokes constrained Stein variational gradient descent to reconstruct fluid flows from limited noisy measurements. These studies have demonstrated the merits of introducing physics constraints. However, technical challenges remain in developing effective physics-informed DL models for super-resolution, especially for irregular domain problems in label-scarce or label-free scenarios. In this work, we developed a physics-informed deep learning framework for super-resolving and denoising LR noisy flow fields with irregular geometries, where the HR data (labels) are not required. The novel contributions of this paper are summarized as follows: (a) we explored a deep learning solution for flow super-resolution without relying on HR data for training given well-posed physics; (b) the proposed method can simultaneously infer unknown conditions if the physics is partially known (e.g., boundary conditions is unknown); (c) we demonstrated the effectiveness of the proposed method on several fluid problems with irregular geometries and non-Gaussian noises in high-dimensional parameter space. The rest of the paper is organized as follows.  The methodology of the proposed method is introduced in Section~\ref{sec:meth}. Numerical results of several test cases, including vascular flow governed with known, unknown, and parametric boundary conditions, are presented in Section~\ref{sec:result}. To illustrate the model's denoising capability, both Gaussian and Non-Gaussian noises (e.g., MRI noise) are studied. Finally, Section~\ref{sec:conclusion} concludes the paper.

\section{Methodology}
\label{sec:meth}

\subsection{Overview}
This work aims to reconstruct a high-resolution flow field from the corresponding low-resolution (possibly noisy) data obtained either by the low-fidelity simulations or measurements. Mathematically, this process can be described by the following mapping,
\begin{equation}
\Scal: (\hat{\Psibold}^l;\mubold)\mapsto(\hat{\Psibold}^h;\mubold),
\end{equation}
where $\hat{\Psibold}^l$ denotes the low-resolution (LR) noisy velocity field on a coarse mesh (e.g., 4D flow MRI measurements), and $\hat{\Psibold}^h$ denotes the high-resolution (HR) noise-free flow field; $\mubold$ represents the vector of physical parameters (e.g., geometry, inflow/outflow boundary conditions, and flow properties). In general, the dimension of $\hat{\Psibold}^h$ is much higher than that of $\hat{\Psibold}^l$ and thus could reveal more details of the flow field. We aim to develop a deep learning (DL) based super-resolution (SR) solution, where a CNN model $\Scal^{c}$ is trained to approximate this LR-to-HR mapping as $\Scal\approx \Scal^{c}$.  Once fully trained, the CNN model can be used to super-resolve any given LR data and generate the corresponding HR noise-reduced flow solutions. In contrast to the previous works, the proposed SR-CNN will be trained purely based on physical laws with strictly imposed boundary conditions and thus does not need any HR data (i.e., labels). Moreover, the proposed learning framework is able to assimilate sparse observation data, unifying the forward and inverse modeling processes. Namely, when the underlying physics are partially known (e.g., boundary condition or other physical parameters are unknown), extra observations can be assimilated to enable forward super-resolution and inverse determination of unknowns simultaneously. 
\begin{figure}[htp!]
	\centering
	\includegraphics[width=1\textwidth]{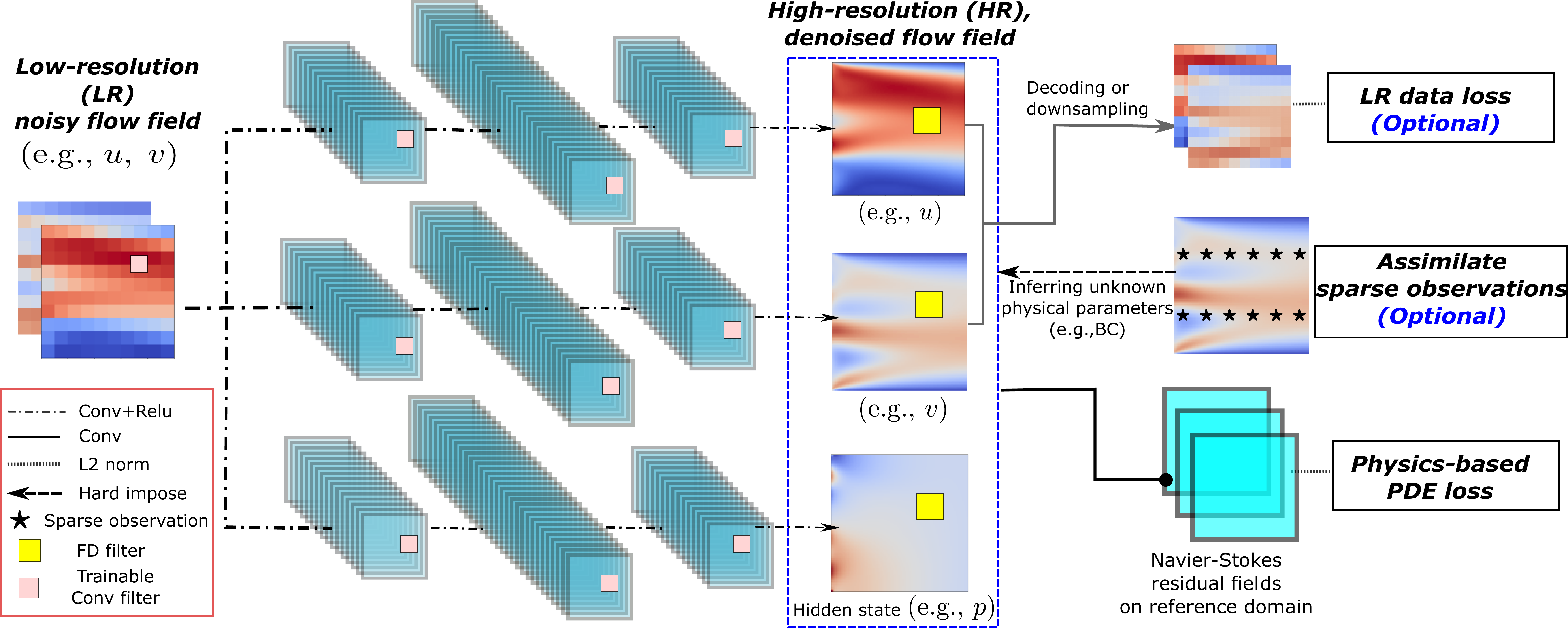}
	\caption{The schematic of physics-informed CNN for flow super-resolution.}
	\label{fig:overall}
\end{figure}
The overall schematic of the proposed physics-informed DL strategy for flow super-resolution is shown in Fig.~\ref{fig:overall}, and each component of the framework will be detailed in the following subsections.
%The overall framework of applying CNN to super-resolve the flow field is drawn in the Fig.~\ref{fig:overall}. The CNN model takes the low-resolution flow field (usually noisy) as the input to perform a serious of convolution operations, then generates the super-resolved, denoised, high-resolution flow field. To train the CNN model (tune CNN parameters), the physics-informed loss function is evaluated instead of the classic data-misfit loss function so that the model can be completely trained without high-resolution labeled data. In case there can be unknown boundary conditions that make PDEs ill-posed, the proposed framework is enabled to assimilate extra observation data to infer the unknowns, yet still to generate the reasonable high-resolution flow field. This section will detail this framework piece by piece.

\subsection{Learning architecture for super-resolution}
A composite DL architecture is constructed (Fig.~\ref{fig:overall}), which takes the (possibly noisy) LR velocity fields as the input channels and produces HR flow solutions. Separate sub-CNN is designed to capture each solution field individually, and thus the trainable network parameters are decoupled for different state variables with different magnitudes. This composite learning structure with decoupled sub-nets has been demonstrated effective in enhancing the learning performance for multivariate regression problems~\cite{SUN2019112732,guo2020ssr,gao2020phygeonet}. The composite DL model consists of several convolutional decoders. Each of them has an identical structure of three hidden convolution layers, which is a classic CNN structure for single image super-resolution (SISR)~\cite{shi2016real}. Specifically, the input layer is first up-sampled to the target resolution using the bicubic interpolation and then goes through three convolution layers with trainable filters of size $5 \times 5$, where 2D convolution operations with the padding of 2 and stride of 1 are applied. All the sub-nets are trained simultaneously with a unified physics-informed loss function as detailed in Section~\ref{sec:pinn}. The trainable parameters of the network are initialized from a uniform distribution of $\mathcal{U}\left(-\sqrt{\frac{1}{25C_{in}}},\sqrt{\frac{1}{25C_{in}}}\right)$, where $C_{in}$ is number of input channels. The network's hyper-parameters are summarized in Table~\ref{tab:CNNset}.
\begin{table}[H]
	\begin{center}
		\begin{tabular}{ |c| c| c| c|}
			\hline
			learning rate &\# of hidden layers& \# of hidden channels& Optimizer\\ 
			\hline
			$10^{-3}$ & 3 & [16,32,16]& Adam\cite{kingma2014adam} \\
			\hline
			padding size&strides&kernel size& Non-lienarity\\ 
			\hline
			2& 1&$5\times5$& ReLU\cite{nair2010rectified}\\
			\hline
		\end{tabular}
	\end{center}\vspace{-1.5em}
	\caption{Hyper-parameters of the sub-CNN.}
	\label{tab:CNNset}
\end{table}

\subsection{Physics-informed training}
\label{sec:pinn}
The composite network $\Scal^{c}$ needs to be trained in order to approximate the LR-to-HR mapping $\Scal$ as,
\begin{equation}
\hat{\Psibold}^h = \Scal(\hat{\Psibold}^l) \approx \Scal^{c}(\hat{\Psibold}^l;\Wbm^c),
\end{equation}
which is an optimization problem traditionally solved based on a large amount of labeled data. Namely, when a set of $n_d^h$ LR/HR data pairs $\{\hat{\Psibold}^l_i, \hat{\Psibold}^h_i\}_{i=1}^{n_d^h}$ are available, the network trainable parameters $\Wbm^c $ can be optimized by iteratively minimizing the mismatch between the CNN predictions $\Scal^{c}(\hat{\Psibold}^l_i)$ and the HR labels $\hat{\Psibold}^h_i$ as follows,
\begin{equation}
\label{eqn:datatraining}
\tilde{\Wbm}^c = \underset{\Wbm^c}{\arg\min} \sum_{i=1}^{n_d^h} \underbrace{\left\| \Scal^{c}(\hat{\Psibold}^l_i;\Wbm^c) - \hat{\Psibold}^h_i \right\|_{\Omega_{p}}}_{\text{data-based loss:}\ \mathcal{L}^d},
\end{equation}
where $\| \cdot \|_{\Omega}$ denotes the $L_2$ norm over the entire domain ${\Omega}_{p}$. However, this \emph{data-driven} training process requires enormous labeled data (i.e., HR samples), which are usually less accessible and way more expensive to acquire than the input data (i.e., LR samples). In many cases, the HR labels are even not available at all due to the resolution limit of the measurement techniques. In this work, we try to tackle this challenge and develop a CNN-SR solution \emph{without relying on HR data} as training labels. A physics-informed learning strategy is adopted to enable label-free training in data-sparse/absent scenarios. The general idea is to leverage the (partially) known physics of the fluid flow (e.g., conservation laws and boundary conditions) to drive the CNN training such that the super-resolved flow information is learned from the flow governing equations instead of massive HR labels. Here we consider fluid problems governed by the steady incompressible Navier-Stokes equations parameterized by $\mubold$, 
\begin{equation}
\label{eq:ns}
\mathscr{R}(\ubm, p; \mubold) = \mathbf{0} := \left \{
\begin{aligned}
&\nabla \cdot \ubm = \mathbf{0},  &\mathrm{in}\ \Omega_{p},\\
&(\ubm\cdot\nabla)\ubm + \frac{1}{\rho}\nabla p - \nu\nabla^2\ubm + \mathbf{b}_f = \mathbf{0}, &\mathrm{in}\ \Omega_{p}, 
\end{aligned} \right .
\end{equation}
where $\ubm$ is the velocity and $p$ is the pressure; $\nu$ and $\mathbf{b}_f$ represents the viscosity and body force of the fluid flow, respectively. The flow solutions can be uniquely determined with given boundary condition (BC), $\mathscr{B}(\ubm, p; \mubold) = 0, \mathrm{on}\ \partial\Omega_p$. Since the CNN super-resolved flow fields should satisfy the governing equations, the training can be recast as a constrained optimization problem by minimizing the PDE residuals,
\begin{equation}
\label{eqn:pdetraining}
\begin{split}
&\tilde{\Wbm}^c = \underset{\Wbm^c}{\arg\min} \sum_{i=1}^{n_d^p} \underbrace{\left\| \mathscr{R} \left( \Scal^{c}(\hat{\Psibold}^l_i;\Wbm^c) \right) \right\|_{\Omega_{p}}}_{\text{PDE-based loss:}\ \mathcal{L}^p},\\
&s.t.\;\; \mathscr{B}\left(\Scal^{c}(\hat{\Psibold}^l_i;\Wbm^c); \mubold\right) = 0, \mathrm{on}\ \partial\Omega_p,
\end{split}
\end{equation}
where $\hat{\Psibold}^l_i = \ubm(\Xboldcal^l; \mubold_i)$ is the LR velocity field discretized on a coarse mesh $\Xboldcal^l$, and $\hat{\Psibold}^h_i = [\ubm(\Xboldcal^h; \mubold_i), p(\Xboldcal^h; \mubold_i)]^T \approx \Scal^{c}(\hat{\Psibold}^l_i;\Wbm^c)$ is the CNN super-resolved flow fields on a fine mesh $\Xboldcal^h$. To evaluate the PDE residuals on the discretized domain, we use convolution operations with the finite difference filters to compute the derivative terms in Eq.~\ref{eq:ns}, and the details are given in \ref{sec:Convolution_Operators_for_Graident_and_Laplacian}. The boundary condition is strictly enforced into the CNN architecture, where the boundary operator $\mathscr{B}$ is discretized and imposed on the CNN-SR predictions in a hard manner using padding operations~\cite{gao2020phygeonet}. In contrast to the traditional label-based data-driven approach, the number $n_d^p$ of training samples is not constrained by the availability of HR data. Hence, the training space can be freely explored with a large number of LR data $\hat{\Psibold}^l_i$, which are assumed to be very cheap to obtain. Moreover, to further facilitate the PDE-based training, the CNN-SR output can be down-sampled via  the pooling operation or an extra encoder) to construct a \emph{LR data loss}, the mismatch between LR input data and downsampled CNN-SR output predictions (as shown in Fig.~\ref{fig:overall}). However, the LR-data loss should only be included if the LR data are noise-free; otherwise, the network may tend to overfit the data noise.

\subsection{Assimilate sparse observation data for partially known physics}
It is commonly known that the flow physics is governed by the Navier-Stokes equations $\mathscr{R}(\ubm, p; \mubold) = 0$, but some of the physical parameters $\mubold$, such as inlet profiles or fluid properties, are unknown in many cases. On the other hand, it is possible to access additional observation data, which, however, is often spatially sparse and/or indirect to the quantity of interest. For example, the 3D full-field velocity information is obtained using 4D flow MRI techniques in cardiovascular applications, but the spatial resolution and SNR are unsatisfied and need enhancement~\cite{ong2015robust}. More accurate flow field data can be observed by 2D phase-contrast MR imaging, which is only available on a limited number of 2D slices. The proposed physics-informed CNN-SR framework can naturally leverage these additional sparse observations to enable both forward super-resolution and inverse parameter determination in a unified manner. Here we introduce a novel approach of assimilating additional sparse observation data to infer the under-determined physical conditions/parameters $\mubold^s \subset \mubold$. First, the unknowns (e.g., boundary conditions) are parameterized as a trainable vector $\mubold^s$, which are incorporated into the SR learning architecture either through the equation-based loss function or strictly-imposed boundary conditions. Second, the sparse observations $\Ybm^{obs}$ are assimilated into the network in a hard manner, where the CNN-SR predictions are strictly enforced to be equal to the data at sparse locations by constructing the model output $\Psibold^c$ as, 
\begin{equation}
\Fcal^{s2o}(\Psibold^c) = \Fcal^{s2o}(\hat{\Psibold}^c)\times 0 + \Ybm^{obs},
\end{equation}
where $\hat{\Psibold}^c$ is the CNN-SR raw output and $\Fcal^{s2o}: \Psibold \to \Ybm$ indicates the state-to-observable map. Therefore, the constrained optimization for PDE-driven training is recast as,
\begin{equation}
\label{eqn:pdetraining-data}
\begin{aligned}
&\tilde{\Wbm}^c, \mubold = \underset{\Wbm^c, \mubold}{\arg\min} \sum_{i=1}^{n_d^p} \underbrace{\left\| \mathscr{R} \left( \Scal^{c}(\hat{\Psibold}^l_i;\Wbm^c); \mubold\right) \right\|_{\Omega_{p}}}_{\text{PDE-based loss:}\ \mathcal{L}^p},\\
&\mathrm{s. t.}
\left \{
\begin{split}
&\mathscr{B}\left(\Scal^{c}(\hat{\Psibold}^l_i;\Wbm^c); \mubold\right) = 0, \mathrm{on}\ \partial\Omega_p,\\
&\Fcal^{s2o}\left(\Scal^{c}(\hat{\Psibold}^l_i;\Wbm^c)\right) - \Ybm^{obs} = 0,
\end{split} \right.
\end{aligned}
\end{equation}
where both the network parameter vector $\tilde{\Wbm}^c$ and physical parameter vector $\mubold$ are inferred simultaneously. It worth mentioning that the proposed ``hard'' DA approach is based on the assumption that the observation data are relatively precise. When assimilating very noisy observations, the penalty-based ``soft'' approach used in the PINN~\cite{raissi2019physics} should be employed to avoid overfitting the data noise. 

\subsection{Coordinate transformation for irregular domain}
A general limitation of CNNs is that they can only handle problems defined on rectangular domains with uniform grids since the convolution operations are originally designed for processing images described on uniform meshes. However, the geometries in most scientific applications are complex and irregular (e.g., subject-specific vessel geometries in cardiovascular applications). In order to perform physics-informed super-resolution on non-rectangular domains, we adopt the geometry-adaptive CNN formation proposed by Gao et al.~\cite{gao2020phygeonet}, where the elliptic coordinate transformation is utilized to reformulate the PDE-constrained learning from the irregular physical domain ($\xbm \in \Omega_p$) to the regular reference domain ($\xibold \in \Omega_r$). Particularly, the one-to-one coordinate transformation map $\mathcal{G}: \Omega_r \to \Omega_p$ is obtained numerically by solving an elliptic problem, e.g., diffusion equations. The Jacobians of the map  $\mathcal{G}$ are then computed to convert differential operators from the physical domain to the reference domain, 
\begin{linenomath*}
	\begin{subequations}
		\label{eqn:Du}
		\begin{alignat}{2}
		\frac{\partial }{\partial x}&= {\frac{1}{J}}\Big[\left(\frac{\partial }{\partial \xi}\right) {\left(\frac{\partial y}{\partial \eta}\right)} - \left(\frac{\partial }{\partial \eta}\right) {\left(\frac{\partial y}{\partial \xi}\right)}\Big],\\
		\frac{\partial }{\partial y}&={\frac{1}{J}}\Big[\left(\frac{\partial }{\partial \eta}\right) {\left(\frac{\partial x}{\partial\xi}\right)} - \left(\frac{\partial }{\partial \xi}\right) {\left(\frac{\partial x}{\partial \eta}\right)}\Big],
		\end{alignat}
	\end{subequations}
\end{linenomath*}
where coordinates of physical domain and reference domain are $\xbm = [x, y]^T$ and $\xibold = [\xi, \eta]^T$, respectively; $J=\frac{\partial x}{\partial\xi}\frac{\partial y}{\partial\eta}-\frac{\partial x}{\partial\eta}\frac{\partial y}{\partial\xi}$ is the determinant of the Jacobian matrix and metrics $\frac{\partial y}{\partial\eta}$, $\frac{\partial y}{\partial\xi}$, $\frac{\partial x}{\partial\eta}$, and $\frac{\partial x}{\partial\xi}$ can be precomputed and remain constant given $\mathcal{G}$. Using elliptic coordinate transformation, the PDE-based loss function is reformulated on the reference domain, and thus the classic CNN backbone can be directly used for irregular geometries. For more details, the reader is referred to~\cite{gao2020phygeonet}.

\section{Result}
\label{sec:result}
\subsection{Overview}
We demonstrate the physics-informed CNN-SR analysis on several internal flow cases relevant to cardiovascular applications. The LR input will be denoised and enhanced to the high-resolution field for both deterministic and parametric scenarios. We first study the flow field in a 2D vascular domain with a deterministic setting, where the governing PDEs and boundary conditions are well defined. Moreover, we also investigate the scenario that the flow physics is partially known, (e.g., the inlet boundary condition is unknown) to demonstrate the CNN-SR solution of unifying forward and inverse problems with additional observation data. Lastly, we present the parametric SR analysis for internal flows with a parameterized inlet velocity profile in a high-dimensional parameter space. 

\subsubsection{Synthesis of low-resolution, noisy data} 
Synthetic LR data are generated from finite volume (FV)-based CFD simulations on coarse meshes. The simulated LR velocity fields are corrupted by artificial measurement noises. Two types of noise models are considered: (1) Gaussian noise and (2) non-Gaussian flow MRI noise. 
\begin{itemize}
	\item \textbf{Gaussian noise model}: The LR velocity field ${\Psibold}^l$ is corrupted by an independent and identically distributed (i.i.d.) Gaussian noise factor $\epsilonbold \sim \Ncal(\mathbf{0}, \Ibm)$ as,
	\begin{equation}
	\hat{\Psibold}^l = {\Psibold}^l \cdot (\Ibm + c \epsilonbold),
	\end{equation}
	where the parameter $c \in [0, 1]$ controls the noise level.

	\item \textbf{Non-Gaussian MRI noise model}: To mimic the LR data obtained from the flow MR imaging, the five-point balanced phase-contrast method~\cite{johnson2010improved} is employed to encode the CFD velocity field into the phase space by $\Sbm = \Fcal(\Psibold^l)$, where $\Sbm$ is a complex matrix of five column vectors. Different levels of complex Gaussian noise are added to the complex data, and the synthetic noisy MR flow field can be then obtained via the inverse five-point map $\Fcal^{-1}(\cdot)$ following the Ref of~\cite{ong2015robust},
	\begin{equation}
	\begin{split}
	&\hat{\Sbm} = \Sbm + \frac{c}{\sqrt{2}}|\Psibold^l|\cdot \Ibm (\epsilonbold_1+ \epsilonbold_2 i),\\
	&\hat{\Psibold}^l = \Fcal^{-1}(\hat{\Sbm}),
	\end{split}
	\end{equation}
	where $i$ is the imaginary unit, $\epsilonbold_1, \epsilonbold_2 \overset{\text{i.i.d}}{\sim}\Ncal(\mathbf{0},\Ibm)$, and $c\in[0, 1]$ controls the noise level. The Gaussian noise imposed in the phase space will become highly non-Gaussian once being mapped back to the physical velocity space,  For more details of the forward and inverse five-point maps $\Fcal(\cdot)$ and $\Fcal^{-1}(\cdot)$, we refer to the Refs of~\cite{johnson2010improved,ong2015robust}.
\end{itemize}

\subsubsection{Cases setup}
To evaluate the CNN-SR performance, we generate high-resolution CFD data ($\hat{\Psibold}^h$) as the reference. Moreover, the upsampled results from the bicubic interpolation will also be computed for comparison. Both the LR and HR CFD simulations are conducted using OpenFOAM~\cite{jasak2007openfoam}, an open-source C++ library for FV simulations. The relative error metric $e$ is defined as,
\begin{equation}
e=\sqrt{\frac{\|\hat{\Psibold}^c-\hat{\Psibold}^h\|_{L_2}}{\|\hat{\Psibold}^h\|_{L_2}}}.
\end{equation}
Specifically, the FV solutions of the steady incompressible Navier-Stokes equations are solved using the Semi-Implicit Method for Pressure Linked Equations (SIMPLE) algorithms~\cite{pletcher2012computational}, where the Rhie and Chow interpolation with collocated grids is adopted to prevent the pressure-velocity decoupling~\cite{rhie1983numerical}. The nonlinear convection term is discretized based on the Gauss theorem with the second-order bounded linear upwind interpolation (i.e., Gauss linearUpwind Scheme in OpenFOAM), and the diffusion term is discretized using the central Gauss linear interpolation with the explicit non-orthogonal correction for surface normal gradients (i.e., Gauss linear corrected). The physics-informed CNN-SR model is implemented in PyTorch \cite{paszke2017automatic}, and training is conducted on an NVIDIA GeForce RTX 2080 Graphics Processing Unit (GPU) card. The training histories for all the test cases are summarized by Fig.~\ref{fig:convergencehistory} in \ref{sec:convergence}. The code and datasets for this work will be available at \url{https://github.com/Jianxun-Wang/PICNNSR} upon publication.

\subsection{Non-parametric super-resolution}
\begin{figure}[htp]
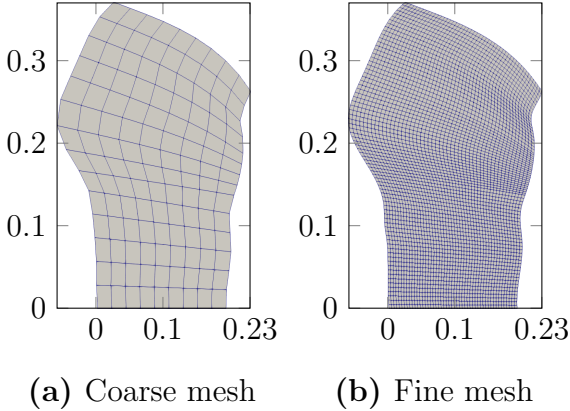

	\centering
	\subfloat[Coarse mesh]
	{\includegraphics[width=0.235\textwidth,height=0.28\textwidth]{case1_coarse_mesh.tikz}}
	\subfloat[Fine mesh]
	{\includegraphics[width=0.235\textwidth,height=0.28\textwidth]{case1_fine_mesh.tikz}}
	\caption{The low-resolution mesh (126 cells) and high-resolution mesh (3773 cells). The LR input data is refined by $30\times$}
	\label{fig:case1mesh}
\end{figure}

The proposed physics-informed CNN-SR model is constructed to super-resolve the LR flow field in a non-parametric setting. Namely, the DL model is trained to learn the LR-to-HR map, $\Scal: (\hat{\Psibold}^l;\mubold)\mapsto(\hat{\Psibold}^h;\mubold)$, where the physical parameter $\mubold$ is fixed. A 2D laminar flow with an irregular vascular geometry is investigated, as shown in Fig.~\ref{fig:case1mesh}. The flow starts at the bottom edge (i.e., inlet) and moves out at the upper edge (i.e., outlet), where the non-slip wall boundary condition is imposed on the left and right boundaries. 
The LR data is obtained from a coarse mesh of $126$ cells, while the HR target mesh has $3773$ cells. The CNN-SR will take the LR data as the input and generates spatially refined data by $30\times$.

%We firstly test the frmework on a deterministic case.We firstly test the frmework on a deterministic case shown in Fig.~\ref{fig:case1mesh}. The low-resolution input mesh has $126$ cells, and the high-resolution mesh has $3773$ cells. We take the velocity field on the coarse mesh calculated by FV as our basic input (added by different noise) and the truth is the velocity filed on the fine mesh.

\subsubsection{Known boundary condition}
\begin{figure}[htp]
\centering
\includegraphics[width=1\textwidth]{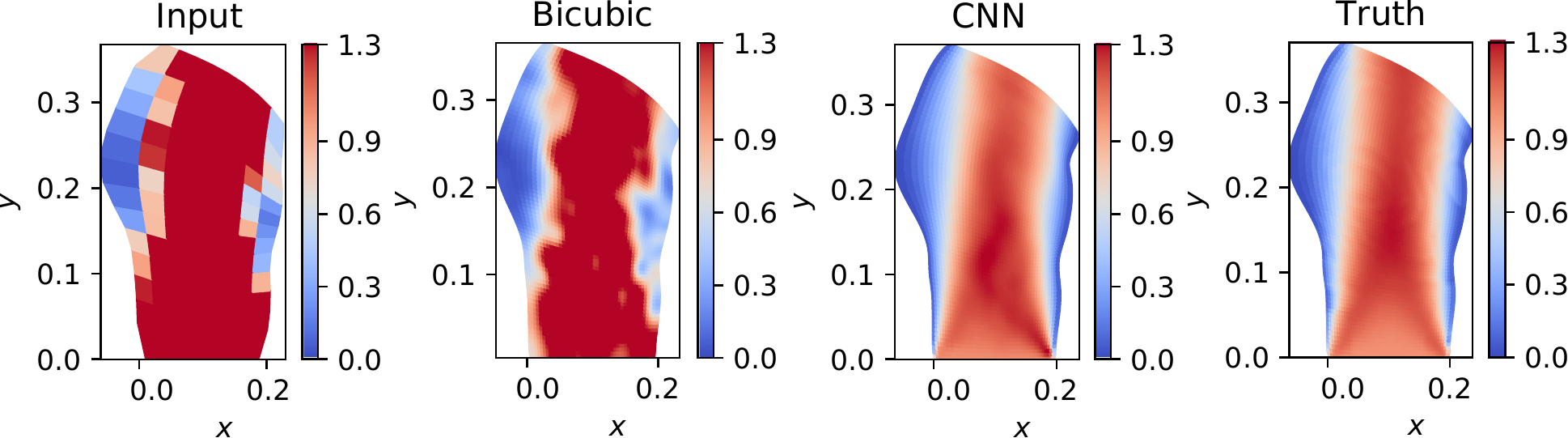}
\caption{The super-resolved results of LR input with the $100\%$ Gaussian noise ($c=1.0$). The relative errors of the bicubic-SR and CNN-SR fields are $0.520$ and $0.067$, respectively.}
\label{fig:deterministic_gaussian}
\end{figure}

We first consider a deterministic scenario with well-posed physics, where both the governing PDEs (i.e., incompressible Navier-Stokes equations) and boundary conditions are well defined. In particular, the inlet boundary condition is known as a constant profile $\ubm = [0, 1]$ and the outlet is defined by $\nabla\ubm\cdot\nbm = 0$ and $p = 0$, where $\nbm$ is the local wall-normal vector. Figure~\ref{fig:deterministic_gaussian} shows the CNN-SR velocity fields from the LR data with a $100\%$ Gaussian noise ($c=1.0$). Due to the mesh coarseness, the LR field presents a mosaic pattern and provides very limited information. The SR solution directly upsampled by the bicubic interpolation is unsatisfactory since the large Gaussian noise makes the bicubic-SR solution highly unphysical. In contrast, the CNN-SR solution well agrees with the HR reference data (truth). The flow details of the boundary layer and velocity development can be accurately captured, where the large Gaussian noises are significantly reduced. The relative error of the CNN-SR field is $0.067$, which is an order lower than that of the bicubic-SR result ($0.520$).
%The boundary condition is known as $\Ubm(y=0)=[0,1]$ on the bottom edge. The the top edge is the outlet. And the side edges are set as non-slip walls. We start with the Gaussian noise.  Figure~\ref{fig:deterministic_gaussian} shows with Gaussian noise added on, the low-resolution velocity field can only provide very vague information. We can roughly tell boundary layer near the wall but fail to see the development of velocity from  the inlet. The improvement from the bicubic interpolation is still very limited. However, the output from the CNN is able to provide a super-resolved velocity field reasonably well. The velocity development from the inlet can be recovered and the boundary layer is smoothly captured.

\begin{figure}[htp]
	\centering
	\includegraphics[width=1\textwidth]{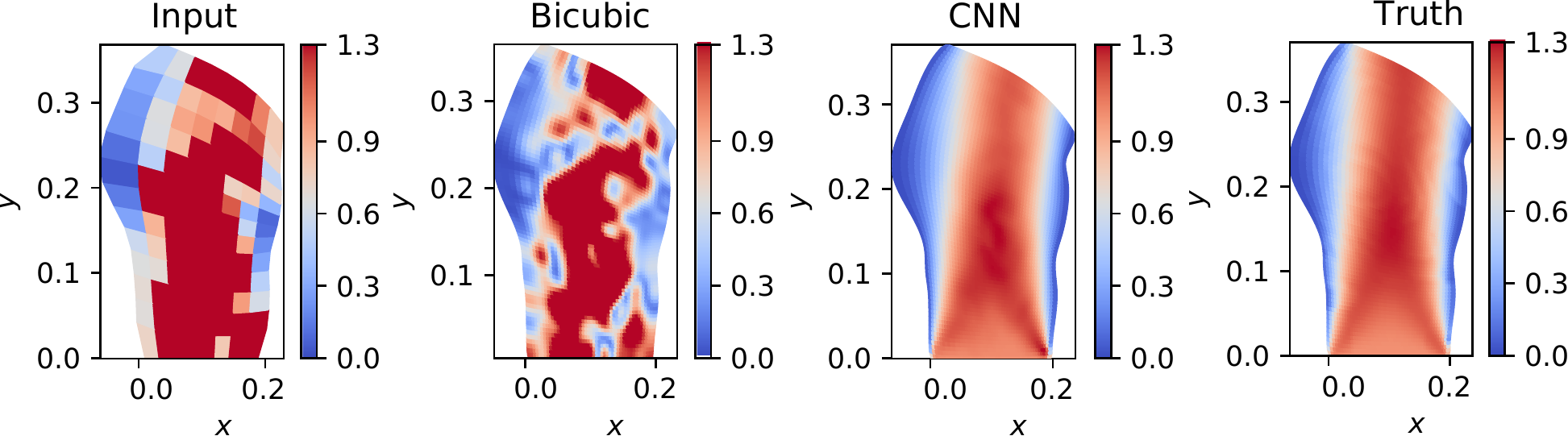}
	\caption{The super-resolved results of LR input with the $100\%$ MRI noise ($c=1.0$). The relative errors of bicubic-SR and CNN-SR fields are $0.617$ and $0.066$, respectively.}
	\label{fig:case1MRI}
\end{figure}

For the experiment with artificial MRI noises shown in Fig.~\ref{fig:case1MRI}, the LR velocity field becomes even more unphysical and the flow information is overwhelmed by the large non-Gaussian MRI noises. Using bicubic interpolation does not show any improvement. Compared to the HR reference, the Bicubic-SR velocity field barely captures any flow physics and has a large relative error of $0.617$. However, the CNN-SR model, informed by the Navier-Stokes equations, can largely remove the MRI noises and generate an accurate super-resolved velocity field with a relative error of $0.066$. The encouraging results show great promise of the proposed method for enhancing the spatial resolution of 4D flow MR imaging, for which the HR labels are often unavailable.
%Then we add MRI noise on the input field. As shown in the Fig.~\ref{fig:case1MRI}, the MRI noise makes the input field nonphysical, the velocity filed displays rapid changes on some places in the domain. And the boundary layer is very difficult to recognize. Even with after bicubic interpolation, the field does not have any improvement on providing flow information. However, CNN can still generate very good super-resolved field.

\subsubsection{Unknown boundary condition}
\begin{figure}[H]
	\centering
	\includegraphics[width=0.9\textwidth]{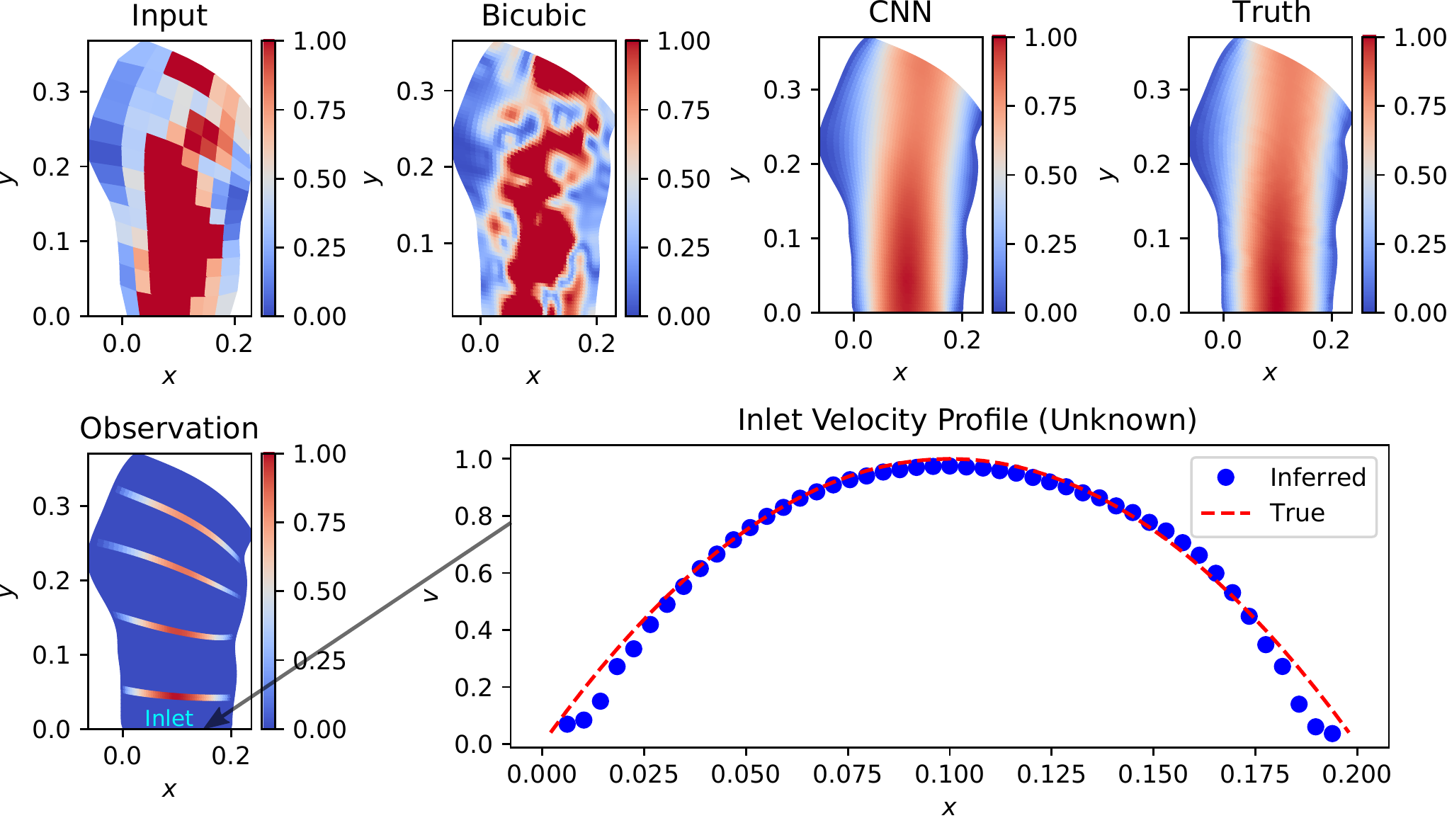}
	\caption{The super-resolved results of LR input with the $100\%$ MRI noise ($c=1.0$) where inlet boundary is unknown. The sparse velocity observations (on $\sim2\%$ of mesh grids) are assimilated to infer the inlet velocity profile (a 45-dimensional field).}
	\label{fig:MRIInverse}
\end{figure}

In this subsection, we demonstrate the capability of unifying the super-resolution and data assimilation for the situation where the physics is ill-posed (e.g., unknown boundary conditions), but additional sparse observation data is available. This scenario is quite common in cardiovascular applications. The 4D flow MR imaging techniques enable noninvasive and \emph{in vivo} measurements of full-field blood flow information, whose spatial resolution, however, is too low to perform any quantitative analysis. Although the Navier-Stokes equations can be used to refine the LR data, the boundary conditions (e.g., inlet velocity field and outlet pressure distributions) are often not available in clinical practice. On the other hand, some sparse high-fidelity observations can be obtained by the 2D PC-MRI on a few 2D slices, which can be assimilated to infer the unknown boundaries. To mimic this scenario, we conduct a numerical experiment with the same setting as above, where the synthetic LR data corrupted by $100\%$ MRI noises are super-resolved using the Navier-Stokes-informed CNN. In contrast to the previous example, we here assume the true inlet velocity profile $\ubm = [0, 100x(0.2-x)]$ is unknown, but more accurate velocity observations are given only on four slices ($\sim2\%$ of mesh grids), as an analogue of the sparse 2D PC-MRI data (see Fig.~\ref{fig:MRIInverse}). It can be seen that not only the LR noisy flow field has been super-resolved to be in a good agreement with the HR reference ($e = 0.029$), but also the unknown inlet velocity profile (a 45-dimensional field) can be accurately recovered in a unified manner.

%Here we show the situation that the boundary condition is unknown but with partial high-resolution observation. This can happen very often in the practical vascular medical treating, the partial high-resolution data (local) comes from phase contrast MRI (PC–MRI). The low-resolution input (global) comes from the 4-D MRI. We set the true inlet boundary condition here as $\Ubm(x)=[0,100*x*(0.2-x)]$, which formulates a quadratic profile. However, we do specify this boundary condition in CNN, and leave CNN to infer it from the sparse high-resolution observation. Figure~\ref{fig:MRIInverse} shows the results: the low-resolution input corrupted by MRI noise is non-physical; the Bicubic interpolation is very limited to further refine the flow field; by assimilating the high-resolution observation, the CNN model super-resolves the flow field reasonably well, the velocity development and boundary layers are clear to display; furthermore, the unkonwn inlet boundary condition is fully recovered together with the internal field.We also test the Gaussian noise, Fig.~\ref{fig:InferGaussianResult} shows the result. 

\subsection{Parametric super-resolution}
\begin{figure}[htp]
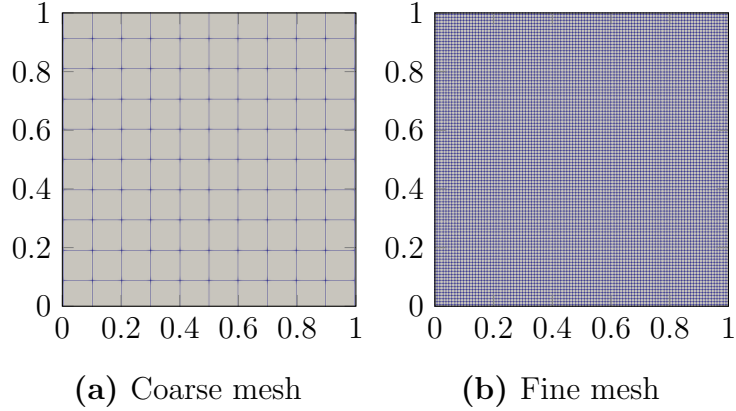

	\centering
	\subfloat[Coarse mesh]
	{\includegraphics[width=0.3\textwidth]{case_parametric_coarsemesh.tikz}}
	\subfloat[Fine mesh]
	{\includegraphics[width=0.3\textwidth]{case_parametric_finemesh.tikz}}
	\caption{The low-resolution input mesh ($10\times10$) and the high-resolution output mesh ($200\times200$). The LR data will be refined by $400\times$}
	\label{fig:case2mesh}
\end{figure}

The proposed CNN-SR solution can be applied for flow super-resolution in the parametric setting, leveraging the powerful interpolatory capability of DL models in high-dimensional parameter space. Namely, the DL model is trained to capture the operator, $\Scal: (\hat{\Psibold}^l;\mubold)\mapsto(\hat{\Psibold}^h;\mubold)$, where the physical parameters $\mubold$ could vary in a high-dimensional space. Once fully trained, the CNN-SR model can be treated as a cost-effective surrogate model that takes the LR data to produce HR solutions. Since the low-fidelity simulations or experiments are relatively cheap to conduct, the trained CNN-SR surrogate could significantly facilitate many query applications, e.g., uncertainty propagation, optimization, and sample-based Bayesian inference. To demonstrate parametric SR capability, we consider internal flows with spatially-varying inflow boundary conditions (including non-zero secondary flow). The CNN-SR is trained to refine the LR flow fields by $400\times$. As shown Fig.~\ref{fig:case2mesh}, the inlet velocity field $\ubold(\xbm)$ is set on the left edge ($x=0$), while the zero-pressure outlet is prescribed on the right edge ($x=1$). Both the top and bottom edges ($y=0,y=1$) are set as non-slip walls. 
Each component of the inlet velocity field ($\ubm = [u(\xbm), v(\xbm)]^T$) is modeled by a scalar stationary Gaussian process,
\begin{equation}
	f(\xbm) {\sim} \Gcal\Pcal\left(\mathbf{0}, K(\xbm,\xbm')\right),\quad K(\xbm,\xbm')=\sigma^2\exp\left(\frac{|\xbm-\xbm'|}{2l^2}\right), \ \ \ \mathrm{in}\  \partial\Omega_{\text{inlet}}
	\label{eqn:gp}
\end{equation}
where $K(\xbm,\xbm')$ is the exponential kernel function; $l$ and $\sigma$ represent the homogeneous length scale and standard deviation of the Gaussian random field. Here we set the length scale $l=0.1$ and the standard deviation $\sigma=0.33$. To represent the Gaussian process in a compact form, we use Karhunen-Loeve (KL) expansion,
\begin{equation}
f(\xbm)=\sum_{i=1}^{n_k\to\infty}\sqrt{\lambda_i}\phi_i(\xbm)\omega_i,
\end{equation}
where $\lambda$ and $\phi(\xbm)$ are eigenvalues and eigenfunctions of the kernel, respectively; $\omega_i$ are uncorrelated random variables with zero mean and unit variance. We further truncate the KL expansion with a finite number ($n_k=10$) of basis to capture $0.96\%$ energy of the random field. The stream-wise velocity field ($u$) and the transverse velocity field ($v$) are defined as,
\begin{equation}
u(\xbm)=1+f(\xbm),\quad v(\xbm)=f(\xbm).
\end{equation}
Hence, the inlet boundary condition is parameterized by a 20-dimensional parameter vector,
\begin{equation}
\mubold=[\omega_1, ..., \omega_{20}]^T \in \mathbb{R}^{20},
\label{eqn:mudistribution}
\end{equation}
where $\omega_1, ...,\omega_{10}$ parameterize the stream-wise velocity and $\omega_{11}, ...,\omega_{20}$ are for the transverse velocity field. The first 10 KL modes are shown in the Fig.~\ref{fig:klmodes}.

\begin{figure}[H]
	\centering
	\includegraphics[width=0.6\textwidth,height=0.05\textwidth]{NS_Parametirc_ColorBar.tikz}
	\vfill
	\includegraphics[width=0.8\textwidth]{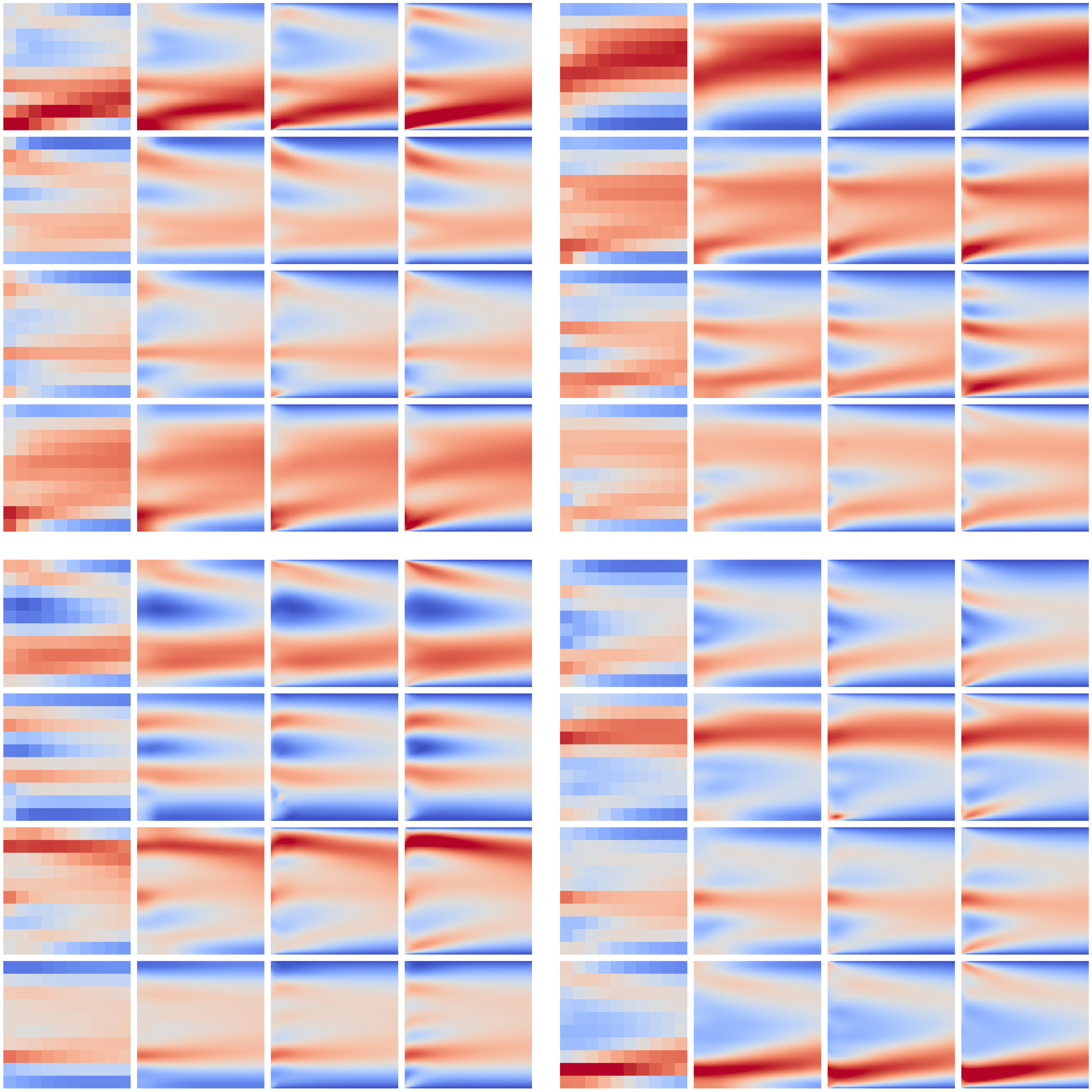}
	\caption{The CNN-SR results of noisy-free LR data on a subset of testing inlet samples randomly drawn from the Gaussian process. In every block, each row has a same inlet velocity field, and the LR data, bicubic-SR results, CNN-SR results, and HR reference are listed from the left to the right.}
	\label{fig:parametric_noise_0_result}
\end{figure}

The CNN-SR model is trained on 15 inlet samples randomly drawn from the Gaussian process defined by Eq.~\ref{eqn:gp}, where the PDE-based loss function is minimized with $10^3$ iterations. Once the CNN-SR model is trained, it can super-resolve the LR flow fields of these 15 inlet samples and be used as a surrogate model to rapidly refine any LR data with unseen inlets in the 20-dimensional parameter space. To evaluate the model's generalizability, we generate $985$ new testing inlets that are unseen during the training. Figure~\ref{fig:parametric_noise_0_result} shows the CNN-SR results of 16 samples randomly selected from the test set, where noise-free LR data are used as the input. Though without any noise, the LR data contain very limited flow information because of the mesh coarseness ($100\times100$). Both bicubic interpolation and trained CNN-SR model can spatially refine the LR data by $400\times$ and show improvements. However, the CNN-SR results reveal more flow details and have a better agreement with the HR reference. 
%We train the CNN model with 15 samples from the multi-variant Gaussian distribution (Eqn.~\ref{eqn:mudistribution}). Then for testing, we sample $985$ more samples to test the performance of the super-resolution via CNN. Figure~\ref{fig:parametric_noise_0_result} shows 16 samples randomly picked up from the testing samples. With noise-free input, the Bicubic interpolation can improve the low-resolution input. But CNN is able to generate a better super-resolved field by resolving more physical flow detials such as boundary layer.

\begin{figure}[H]
	\centering
	\includegraphics[width=0.6\textwidth,height=0.05\textwidth]{NS_Parametirc_ColorBar.tikz}
	\vfill
	\includegraphics[width=0.8\textwidth]{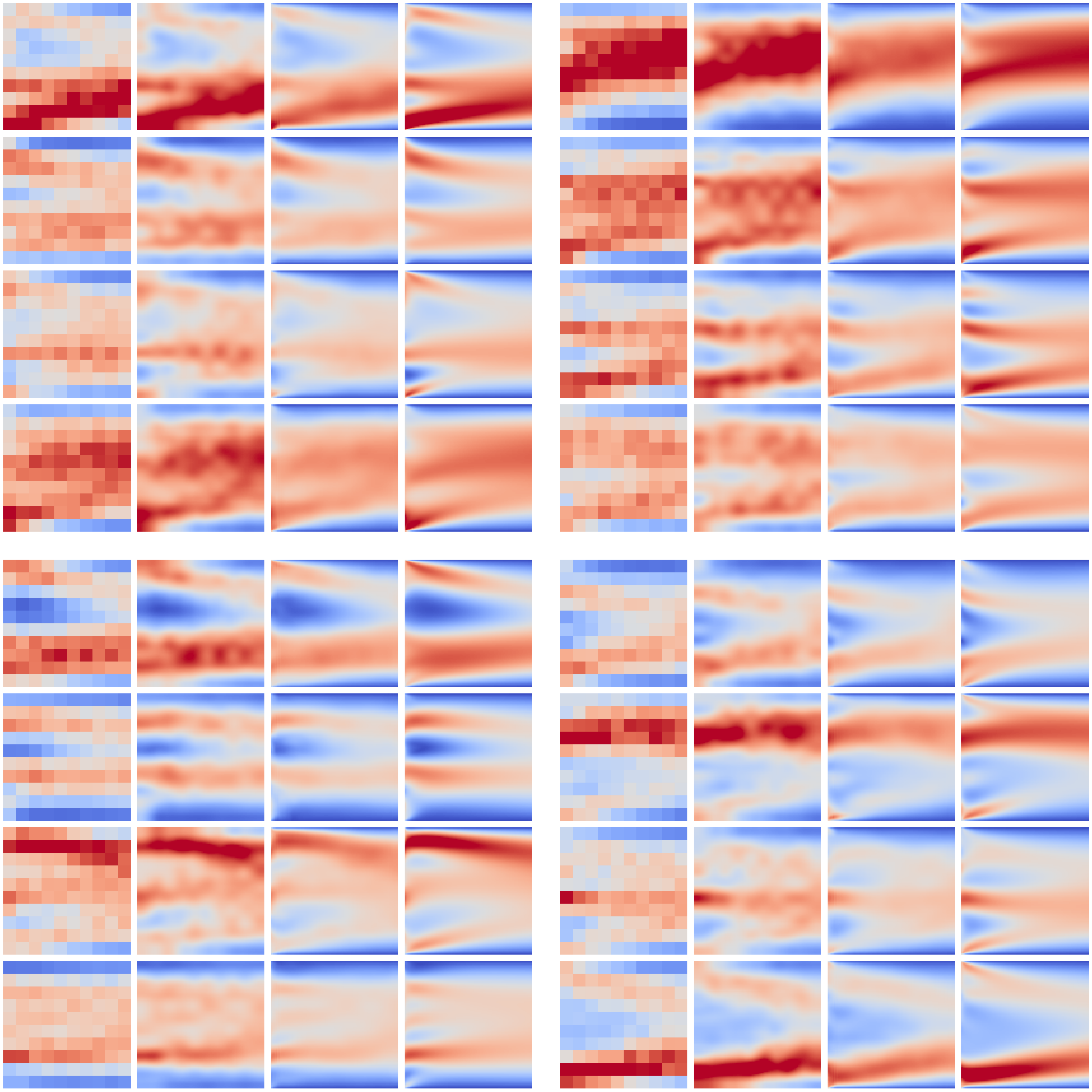}
	\caption{The CNN-SR results of noisy LR data on a subset of testing inlet samples randomly drawn from the Gaussian process. In every block, each row has a same inlet velocity field, and the LR data, bicubic-SR results, CNN-SR results, and HR reference are listed from the left to the right.}
	\label{fig:parametric_noise_20_result}
\end{figure}

\begin{figure}[H]
	\includegraphics[width=1\textwidth,height=0.3\textwidth]{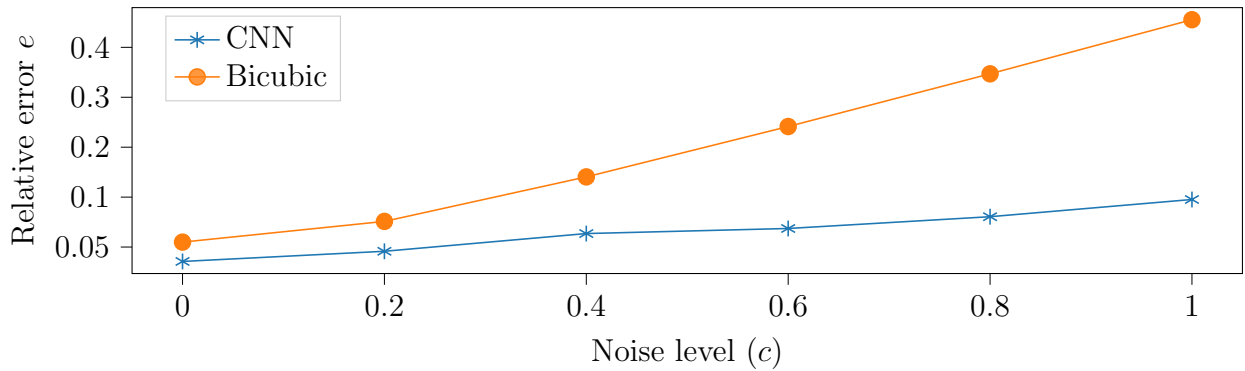}
	\caption{The relative errors of CNN and Bicubic SR results of LR inputs with different noise levels over 1000 testing samples.}
	\label{fig:errortrend}
\end{figure}

When the LR data contain noises, the superiority of the CNN-SR solution becomes more significant compared to the bicubic interpolation. To demonstrate this merit, we conduct another numerical experiment, where the LR data are corrupted with $20\%$ Gaussian noise (noise level $c = 0.2$). The CNN-SR model is trained on the LR data by resampling the noise at every iteration to recognize the noisy inputs. Figure~\ref{fig:parametric_noise_20_result} shows the super-resolution results of noisy LR data. It is apparent that the bicubic-SR velocity fields are visually unphysical by directly interpolating the data noises. The SR performance of the physics-informed CNN model still remains excellent as it accurately refined the spatial resolution of the LR data by $400\times$, and the SR results agree with the HR reference very well on all testing samples.
Figure~\ref{fig:errortrend} shows the mean relative errors versus LR data noise levels for both CNN-SR and bicubic-SR results over 1000 testing samples. We can see the SR performance of bicubic interpolation remarkably deteriorates as the input noise level increases. When the LR data contain $100\%$ noises, the naive bicubic interpolation completely fails as the relative error grows up to nearly $50\%$. In comparison, CNN-SR solutions are less sensitive to the growth of data noise. Although showing a similar trend, the CNN-SR solutions' relative error remains small (less than $10\%$), even the data noise level reaches to $100\%$.

%To make the CNN model handle the noisy (Gaussian noise) input, we re-train the model by adding random noise re-sampled to a same input at each iteration. So the CNN model is trained to be able to filter noise to some extent. Here the noise dimension is $2\times100$ (the same as the degree of freedom of the input). We set $c=0.2$ to control the noise level, and the result is shown in Fig.~\ref{fig:parametric_noise_20_result}. The CNN can super-resolve the noisy low-resolution significantly better than the Bicubic interpolation. However, as the noise level increasing, the performance of CNN and Bicubic both degrade. Still, shown in Fig.~\ref{fig:errortrend}, the CNN outperforms the Bicubic at different noise level with a smaller error-noise slope.

The computational costs of a single model evaluation for the HR CFD and CNN-SR models are listed in Table~\ref{tab:wallclocktime}. For a single run on each parameter point, the speedup of the CNN-SR model is more than $3000$ times compared to the CFD simulation. It shows the potential of using the CNN-SR model as a surrogate for massive queries in the high-dimensional input space, which could enable or facilitate ensemble-based uncertainty quantification or inverse optimization.
\begin{table}[H]
	\begin{center}
		\begin{tabular}{| c|c| c|}
			\hline
			&HR CFD & CNN\\ 
			\hline
		Wall-clock time (s)	&$4$ & $1.189\times 10^{-3}$ \\
			\hline
			Hardware &Intel Xeon(R) Gold 6138 & GeForce RTX 2080\\ 
			\hline
			Speedup&\multicolumn{2}{c|}{3364}\\
			\hline
		\end{tabular}
	\end{center}
	\caption{Computational costs of online evaulation of the HR CFD and CNN-SR models.}
	\label{tab:wallclocktime}
\end{table}

%\begin{table}[H]
%	\begin{center}
%		\begin{tabular}{| c|c| c| c|}
%			\hline
%			&FV & CNN& Bicubic\\ 
%			\hline
%		Wall-clock time (s)	&$4$ & $1.189\times 10^{-3}$ & $7.54\times 10^{-5}$\\
%			\hline
%			Hardware &Intel Xeon(R) Gold 6138 & GeForce RTX 2080& GeForce RTX 2080\\ 
%			\hline
%		\end{tabular}
%	\end{center}
%	\caption{Wall-clock time for both models to super-resolve a low-resolution input.}
%	\label{tab:wallclocktime}
%\end{table}
%
%\todo[inline]{Rewrite the following analysis}
%After the CNN model is trained in offline training phase, the time cost for the CNN and the OpenFOAM to perform a singe run of super-resolution is listed in Table~\ref{tab:wallclocktime}. Compared with the finite-volume method, the speed up is more than $3000$ times. \textbf{The training cost depends on the way to deploy the input training data. If the input data is fully loaded in the GPU memory, the speedup of training cost can be reached. However we take another way that the input data is all loaded in the hard drive, and partial input data is only loaded temporarily in the GPU memory when it is needed for training. In such way, the training cost is $7.446$ single-GPU hour. }

\section{Conclusion}
\label{sec:conclusion}
In this paper, we proposed a novel physics-informed deep learning solution for the spatial super-resolution of flow fields. Leveraging the physical laws and boundary conditions of fluid flows, the training of the CNN-SR model only needs LR samples instead of its HR counterparts as labels. Once sufficiently trained, the CNN-SR model can produce the spatially refined flow field, given a noisy LR input in the parameter space. When the flow boundary conditions are unknown, the proposed framework can naturally assimilate additional sparse observation data to simultaneously enable forward SR and inverse determination of unknown boundary conditions. The effectiveness and merit of the proposed CNN-SR model have been demonstrated on a number of non-parametric and parametric spatial flow SR problems relevant to cardiovascular applications, where both Gaussian and non-Gaussian MRI noises are investigated. In particular, we demonstrated that the CNN-SR model, by training on only 15 LR input samples, is able to accurately refine the spatial resolution by $400\times$ for the flow fields with any new inlet BCs sampled in the 20-dimensional parameter space ($\mubold \in \mathbb{R}^{20}$). Compared to the standard SR approach based on the bicubic interpolation, the CNN-SR model shows significantly higher accuracy and robustness. Compared to the standard FV simulation, the single sample speedup is more than $10^3$ times, showing its potential for many-query applications. The current work is only focused on the spatial super resolution for 2D flow fields. Future work will extend the framework for both spatial and temporal super-resolution of unsteady fluid flows in 3D complex domains.  

%In this paper, we have proposed a novel deep learning framework to perform spatial super-resolution for flow fields. Thanks to the physics-informed loss function, the CNN model only takes the spatial low-resolution without its high-resolution counterpart as labels for training. After the model is finished training, it can generate the high-resolution flow field, given a noisy low-resolution input. 
%In the way of hard imposing the sparse high-resolution data, with unknown boundary conditions, the CNN model can assimilate the observation to perform the super-resolution task and infer the unknown boundary conditions meanwhile. 
%Compared to the standard FV solver, the prediction speed of CNN is over $10^3$ times faster, which shows its potential. Moreover, the CNN model can assimilate data significantly more conveniently than a standard numerical solver.   The future work will focus on improving this method to solve the chaotic (turbulent) flow and the flow in 3-D complex domains (e.g., patient-specific vessels). And we will extend the model to quantify the uncertainty when sparse high-resolution observations are not available at all.    

\section*{Compliance with Ethical Standards}
The authors declare that they have no conflict of interest.

\section*{Acknowledgement}
The author gratefully acknowledge the funds from National Science Foundation (NSF contract: CMMI-1934300) in supporting this project.

\appendix
\section{Convolution operators for gradient and Laplacian terms}
\label{sec:Convolution_Operators_for_Graident_and_Laplacian}
The convolution filters for gradient and laplacian terms are stored as 4D tensors shown as below,
\begin{equation}
\frac{\partial u}{\partial x}\;\mathrm{filter}=
\begin{bmatrix}
\begin{bmatrix}
\begin{bmatrix}
\begin{bmatrix}
0 & 0 & 0 & 0 & 0\\
0 & 0 & 0 & 0 & 0\\
1 & -8 & 0 & 8 & -1\\
0 & 0 & 0 & 0 & 0\\
0 & 0 & 0 & 0 & 0
\end{bmatrix}
\end{bmatrix}
\end{bmatrix}
\end{bmatrix}\times\frac{1}{12\delta x},
\end{equation}
\begin{equation}
\frac{\partial u}{\partial y}\;\mathrm{filter}=
\begin{bmatrix}
\begin{bmatrix}
\begin{bmatrix}
\begin{bmatrix}
0 & 0 & 1 & 0 & 0\\
0 & 0 & -8 & 0 & 0\\
0 & 0 & 0 & 0 & 0\\
0 & 0 & 8 & 0 & 0\\
0 & 0 & -1 & 0 & 0
\end{bmatrix}
\end{bmatrix}
\end{bmatrix}
\end{bmatrix}\times\frac{1}{12\delta y},
\end{equation}
and
\begin{equation}
(\frac{\partial^2 u}{\partial x^2}+\frac{\partial^2 u}{\partial y^2})\;\mathrm{filter}=
\begin{bmatrix}
\begin{bmatrix}
\begin{bmatrix}
\begin{bmatrix}
0 & 0 & -1 & 0 & 0\\
0 & 0 & 16 & 0 & 0\\
-1 & 16 & -60 & 16 & -1\\
0 & 0 & 16 & 0 & 0\\
0 & 0 & -1 & 0 & 0
\end{bmatrix}
\end{bmatrix}
\end{bmatrix}
\end{bmatrix}\times\frac{1}{12\delta x\delta y},\;\;\delta x = \delta y
\end{equation}
%Figure~\ref{fig:FilterPic} shows the visualization of the convolution filters mentioned above.
%\begin{figure}[H]
%	\centering
%	\subfloat[$\frac{\partial u}{\partial x}$ filter]
%	{\includegraphics[height=0.2\textwidth]{./dudx.pdf}}
%	\subfloat[$\frac{\partial u}{\partial y}$ filter]
%	{\includegraphics[height=0.2\textwidth]{./dudy.pdf}}
%	\subfloat[$\frac{\partial^2 u}{\partial x^2} + \frac{\partial^2 u}{\partial y^2}$ filter]
%	{\includegraphics[height=0.2\textwidth]{./laplaceU.pdf}}
%	\caption{Visualization of convolution operators for approximated PDE operators}
%	\label{fig:FilterPic}
%\end{figure}

\section{The KL modes for the spatially-varying inlets}
\begin{figure}[H]
	\centering
	\includegraphics[width=0.9\textwidth]{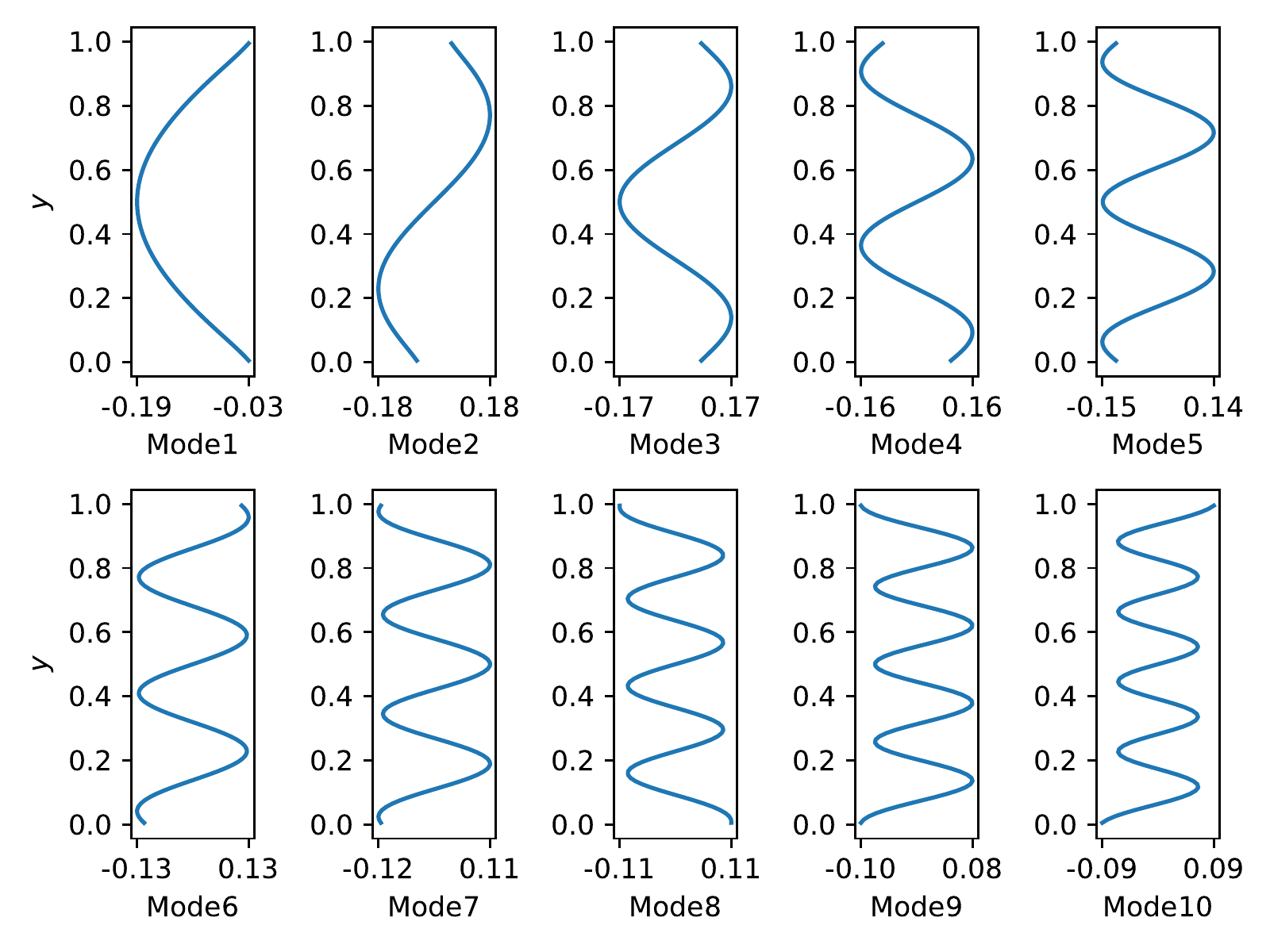}
	\caption{The first 10 KL modes of the Guassian random fields for the spatially-varying inlet boundary conditions.}
	\label{fig:klmodes}
\end{figure}

%\section{Super-resolve with unknown inlet BC}
%\begin{figure}[H]
%	\centering
%	\includegraphics[width=0.8\textwidth]{NS_100Gaussian_Inverese.pdf}
%	\caption{The relative errors of super-resolved velocity magnitude and inlet velocity field is $0.032$ and $0.0706$, respectively. The dashed line is the true inlet velocity profile (45-dimensional field), the cross mark denotes the inferred the inlet velocity profile.}
%	\label{fig:InferGaussianResult}
%\end{figure}

\section{Physics-informed training history}
\label{sec:convergence}
\begin{figure}[H]
	\centering
	{
	\subfloat[{\footnotesize Nonparametric (BC known, Gauss)}]
	{\includegraphics[height=0.3\textwidth]{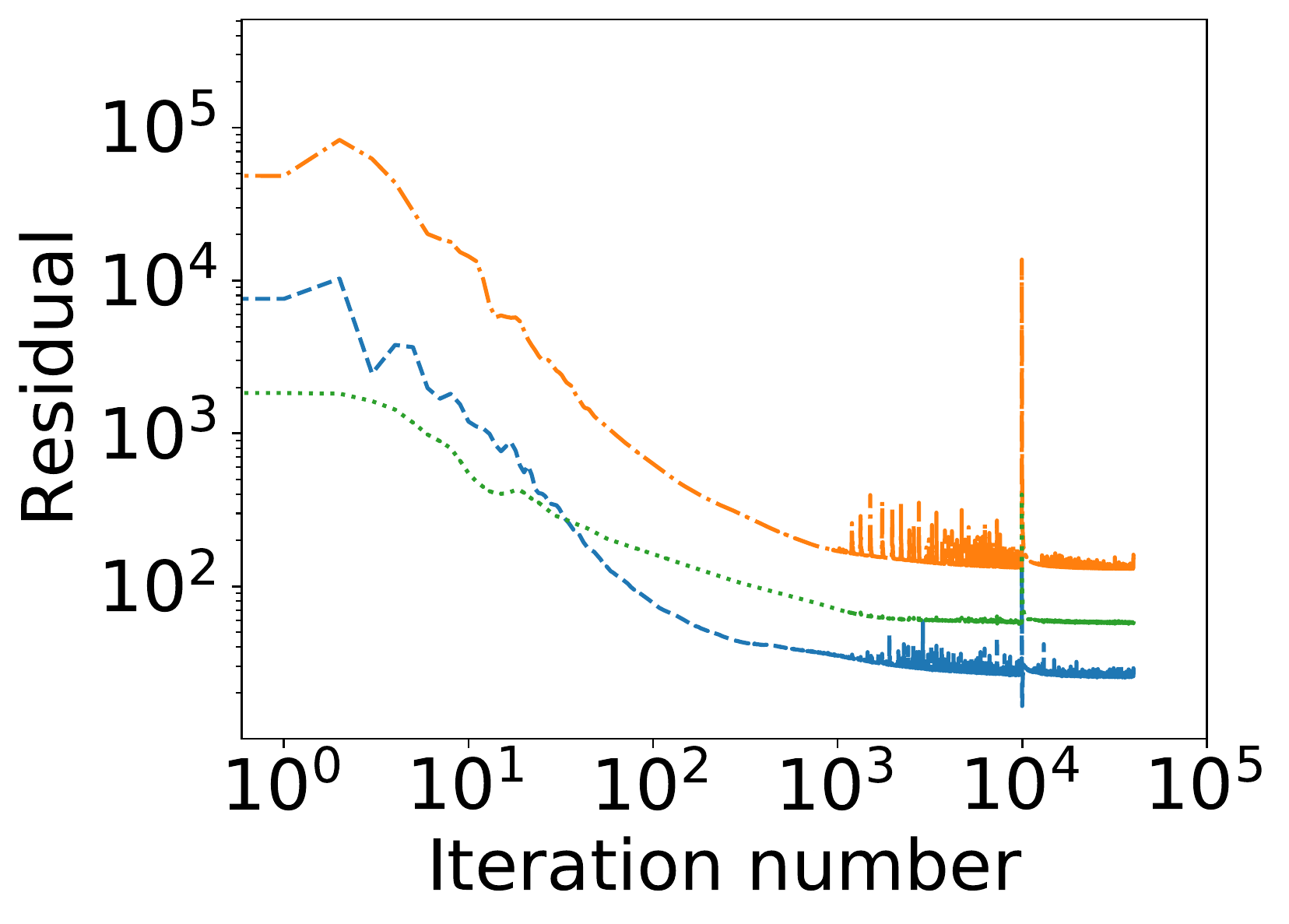}}
	\subfloat[{\footnotesize Nonparametric (BC known, MRI)}]
	{\includegraphics[height=0.3\textwidth]{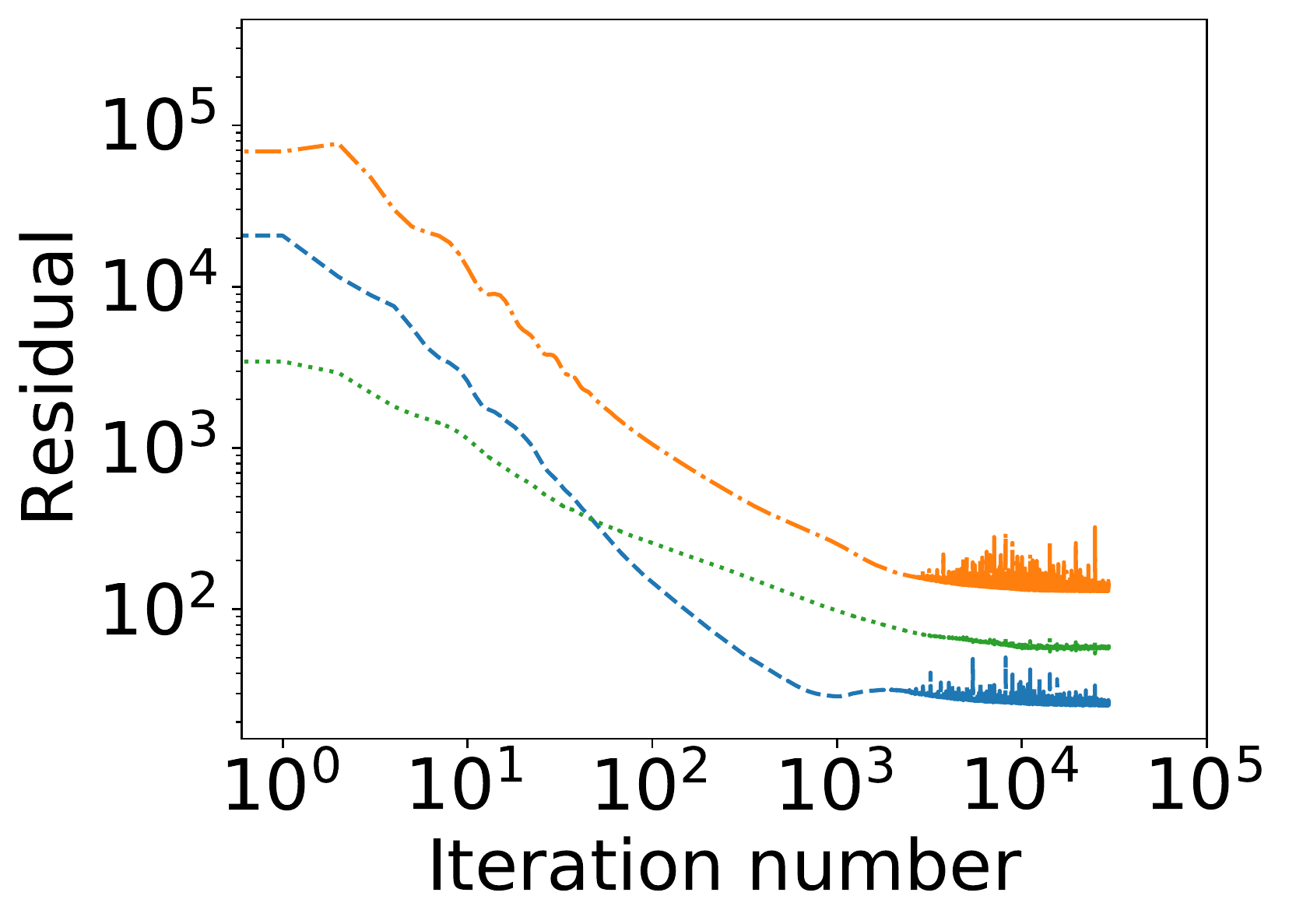}}
	\vfill
	\subfloat[{\footnotesize Nonparametric (BC unknown, Gauss)}]
	{\includegraphics[height=0.3\textwidth]{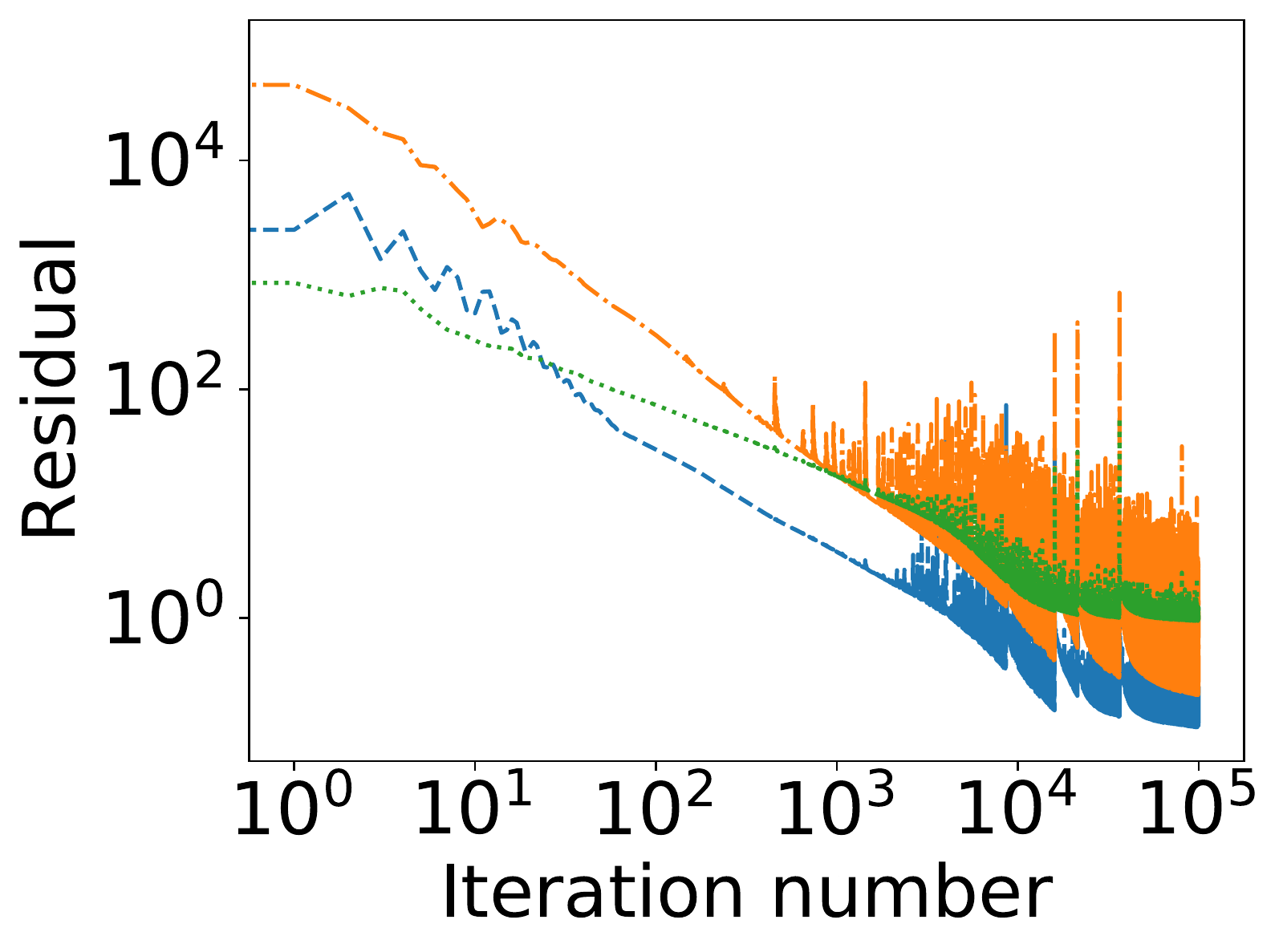}}
	\subfloat[{\footnotesize Nonparametric (BC unknown, MRI)}]
	{\includegraphics[height=0.3\textwidth]{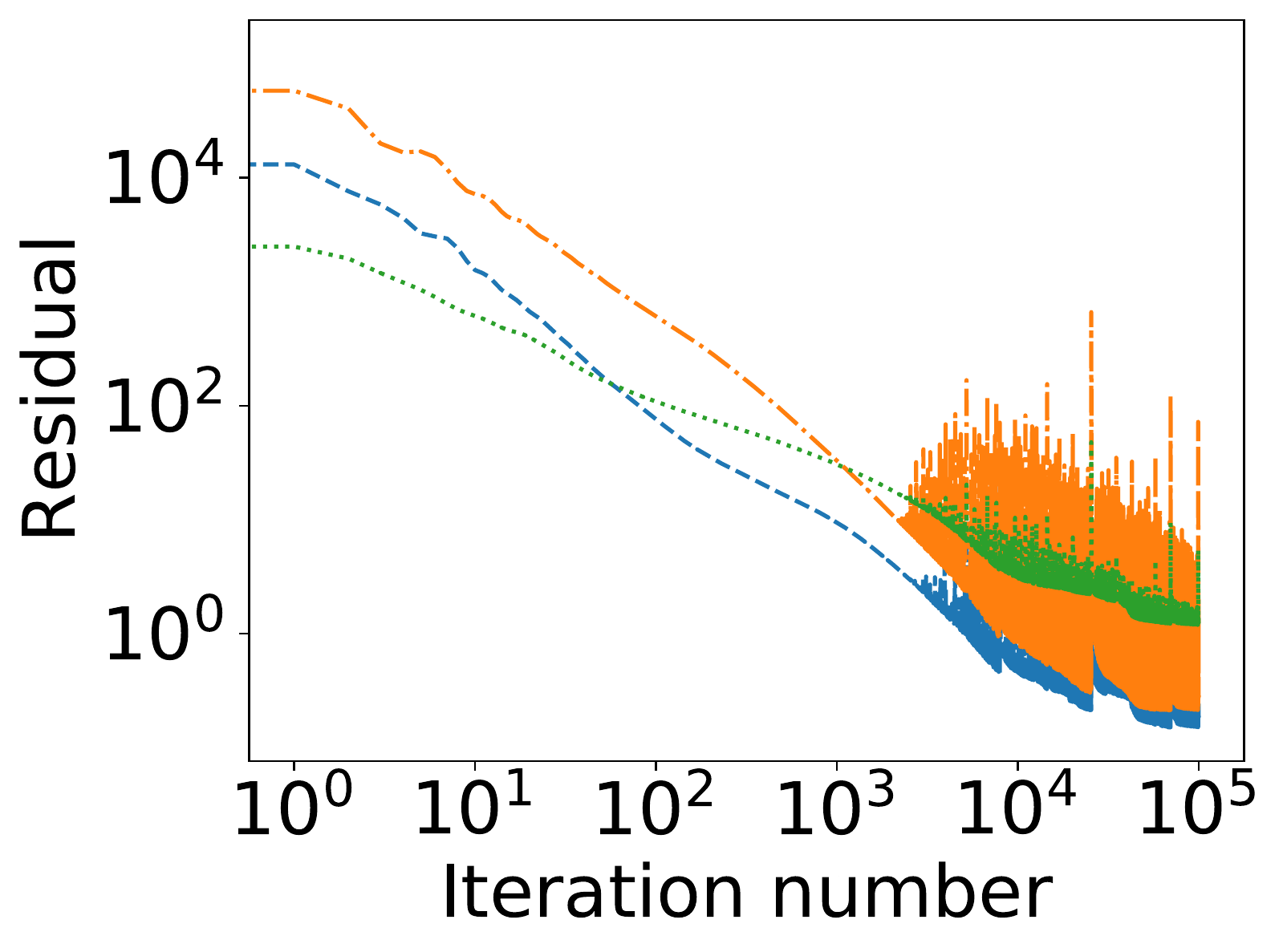}}
	\vfill
	\subfloat[{\footnotesize Parametric (BC known, noise-free)}]{\includegraphics[height=0.3\textwidth]{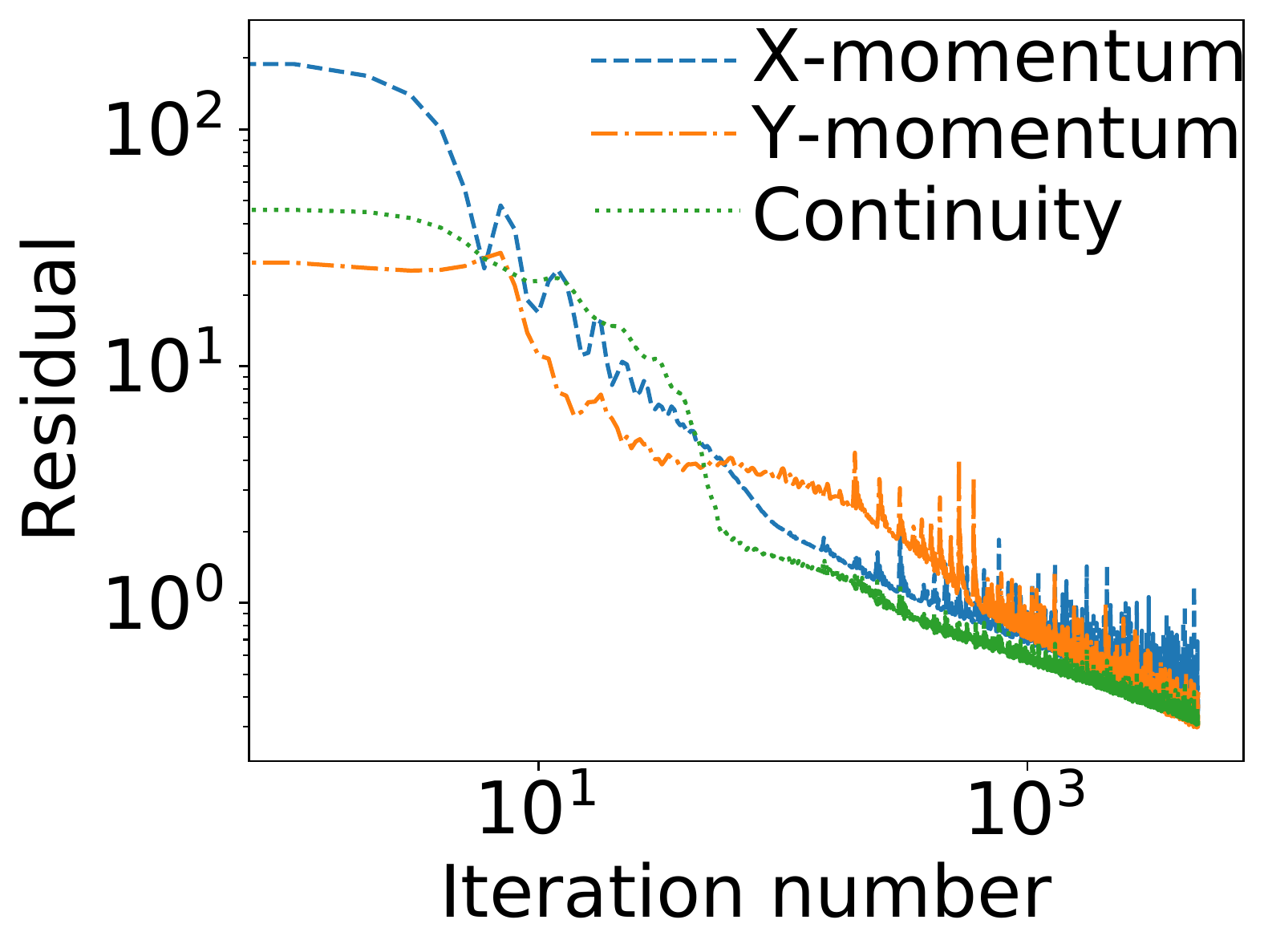}}
	\subfloat[{\footnotesize Parametric (BC known, Gauss)}]{\includegraphics[height=0.3\textwidth]{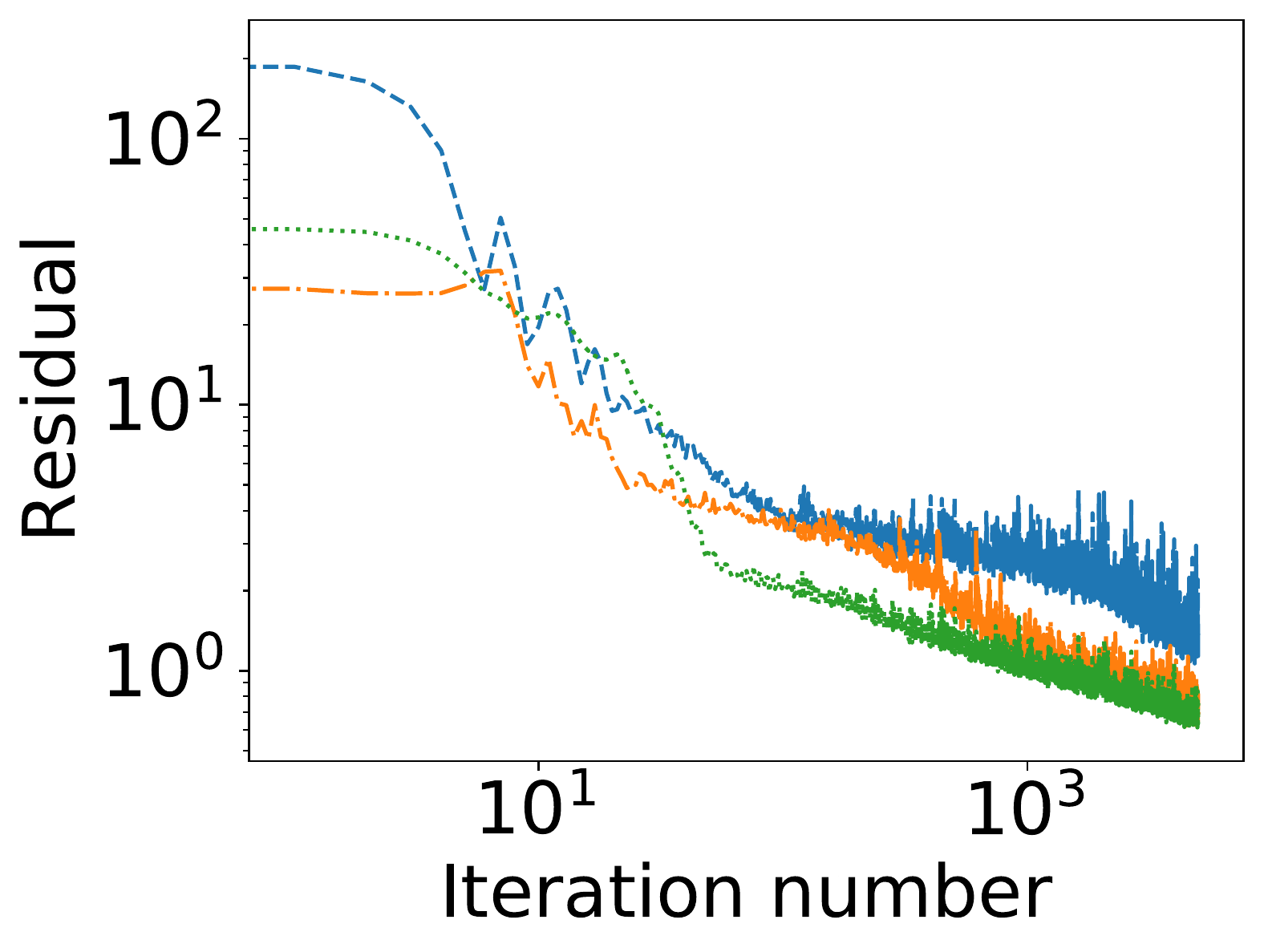}}
}
	\caption{Training histories for nonparametric and parametric super-resolution cases. Each iteration takes about 2 seconds on the NVIDIA 2080 GPU.}
	\label{fig:convergencehistory}
\end{figure}

%\bibliographystyle{elsarticle-num}
%\bibliography{ref}

\begin{thebibliography}{10}
	\expandafter\ifx\csname url\endcsname\relax
	\def\url#1{\texttt{#1}}\fi
	\expandafter\ifx\csname urlprefix\endcsname\relax\def\urlprefix{URL }\fi
	\expandafter\ifx\csname href\endcsname\relax
	\def\href#1#2{#2} \def\path#1{#1}\fi
	
	\bibitem{pollard2016whither}
	A.~Pollard, L.~Castillo, L.~Danaila, M.~Glauser, Whither turbulence and big
	data in the 21st century?, Springer, 2016.
	
	\bibitem{lawley20184d}
	C.~M. Lawley, K.~M. Broadhouse, F.~M. Callaghan, D.~S. Winlaw, G.~A. Figtree,
	S.~M. Grieve, 4d flow magnetic resonance imaging: role in pediatric
	congenital heart disease, Asian Cardiovascular and Thoracic Annals 26~(1)
	(2018) 28--37.
	
	\bibitem{stankovic20144d}
	Z.~Stankovic, B.~D. Allen, J.~Garcia, K.~B. Jarvis, M.~Markl, 4d flow imaging
	with mri, Cardiovascular diagnosis and therapy 4~(2) (2014) 173.
	
	\bibitem{ong2015robust}
	F.~Ong, M.~Uecker, U.~Tariq, A.~Hsiao, M.~T. Alley, S.~S. Vasanawala,
	M.~Lustig, Robust 4d flow denoising using divergence-free wavelet transform,
	Magnetic resonance in medicine 73~(2) (2015) 828--842.
	
	\bibitem{Fathi:2018fv}
	M.~F. Fathi, A.~Bakhshinejad, A.~Baghaie, D.~Saloner, R.~H. Sacho, V.~L. Rayz,
	R.~M. D{\textquoteright}Souza, {Denoising and Spatial Resolution Enhancement
		of 4D Flow MRI Using Proper Orthogonal Decomposition and Lasso
		Regularization}, Computerized Medical Imaging and Graphics 70 (2018) 1--20.
	
	\bibitem{Callaghan:2017jm}
	F.~M. Callaghan, S.~M. Grieve, {Spatial resolution and velocity field
		improvement of 4D-flow MRI}, Magnetic Resonance in Medicine 78~(5) (2017)
	1959--1968.
	
	\bibitem{venturi2004gappy}
	D.~Venturi, G.~E. Karniadakis, Gappy data and reconstruction procedures for
	flow past a cylinder, Journal of Fluid Mechanics 519 (2004) 315--336.
	
	\bibitem{bui2003proper}
	T.~Bui-Thanh, M.~Damodaran, K.~Willcox, Proper orthogonal decomposition
	extensions for parametric applications in compressible aerodynamics, in: 21st
	AIAA Applied Aerodynamics Conference, 2003, p. 4213.
	
	\bibitem{podvin2006reconstruction}
	B.~Podvin, Y.~Fraigneau, F.~Lusseyran, P.~Gougat, A reconstruction method for
	the flow past an open cavity, Journal of fluids engineering 128~(3) (2006)
	531--540.
	
	\bibitem{yakhot2007reconstruction}
	A.~Yakhot, T.~Anor, G.~E. Karniadakis, A reconstruction method for gappy and
	noisy arterial flow data, IEEE transactions on medical imaging 26~(12) (2007)
	1681--1697.
	
	\bibitem{moreno2016aerodynamic}
	A.~I. Moreno, A.~A. Jarzabek, J.~M. Perales, J.~M. Vega, Aerodynamic database
	reconstruction via gappy high order singular value decomposition, Aerospace
	Science and Technology 52 (2016) 115--128.
	
	\bibitem{mifsud2019fusing}
	M.~Mifsud, A.~Vendl, L.-U. Hansen, S.~G{\"o}rtz, Fusing wind-tunnel
	measurements and cfd data using constrained gappy proper orthogonal
	decomposition, Aerospace Science and Technology 86 (2019) 312--326.
	
	\bibitem{schmid2010dynamic}
	P.~J. Schmid, Dynamic mode decomposition of numerical and experimental data,
	Journal of fluid mechanics 656 (2010) 5--28.
	
	\bibitem{tu2013dynamic}
	J.~H. Tu, C.~W. Rowley, D.~M. Luchtenburg, S.~L. Brunton, J.~N. Kutz", On
	dynamic mode decomposition: Theory and applications, Journal of Computational
	Dynamics 1 (2014) 391.
	
	\bibitem{callaham2019robust}
	J.~L. Callaham, K.~Maeda, S.~L. Brunton, Robust flow reconstruction from
	limited measurements via sparse representation, Physical Review Fluids 4~(10)
	(2019) 103907.
	
	\bibitem{manohar2018data}
	K.~Manohar, B.~W. Brunton, J.~N. Kutz, S.~L. Brunton, Data-driven sparse sensor
	placement for reconstruction: Demonstrating the benefits of exploiting known
	patterns, IEEE Control Systems Magazine 38~(3) (2018) 63--86.
	
	\bibitem{foures2014data}
	D.~P. Foures, N.~Dovetta, D.~Sipp, P.~J. Schmid, A data-assimilation method for
	{Reynolds-averaged Navier--Stokes-driven} mean flow reconstruction, Journal
	of Fluid Mechanics 759 (2014) 404--431.
	
	\bibitem{combes2015particle}
	B.~Comb{\`e}s, D.~Heitz, A.~Guibert, E.~M{\'e}min, A particle filter to
	reconstruct a free-surface flow from a depth camera, Fluid Dynamics Research
	47~(5) (2015) 051404.
	
	\bibitem{kikuchi2015assessment}
	R.~Kikuchi, T.~Misaka, S.~Obayashi, Assessment of probability density function
	based on {POD} reduced-order model for ensemble-based data assimilation,
	Fluid Dynamics Research 47~(5) (2015) 051403.
	
	\bibitem{mons2016reconstruction}
	V.~Mons, J.-C. Chassaing, T.~Gomez, P.~Sagaut, Reconstruction of unsteady
	viscous flows using data assimilation schemes, Journal of Computational
	Physics 316 (2016) 255--280.
	
	\bibitem{symon2017data}
	S.~Symon, N.~Dovetta, B.~J. McKeon, D.~Sipp, P.~J. Schmid, Data assimilation of
	mean velocity from 2d piv measurements of flow over an idealized airfoil,
	Experiments in fluids 58~(5) (2017) 61.
	
	\bibitem{wang2016data}
	J.-X. Wang, H.~Xiao, Data-driven cfd modeling of turbulent flows through
	complex structures, International Journal of Heat and Fluid Flow 62 (2016)
	138--149.
	
	\bibitem{xiao2016quantifying}
	H.~Xiao, J.-L. Wu, J.-X. Wang, R.~Sun, C.~Roy, Quantifying and reducing
	model-form uncertainties in reynolds-averaged navier--stokes simulations: A
	data-driven, physics-informed bayesian approach, Journal of Computational
	Physics 324 (2016) 115--136.
	
	\bibitem{brunton2020machine}
	S.~L. Brunton, B.~R. Noack, P.~Koumoutsakos, Machine learning for fluid
	mechanics, Annual Review of Fluid Mechanics 52 (2020) 477--508.
	
	\bibitem{brenner2019perspective}
	M.~Brenner, J.~Eldredge, J.~Freund, Perspective on machine learning for
	advancing fluid mechanics, Physical Review Fluids 4~(10) (2019) 100501.
	
	\bibitem{ling2016reynolds}
	J.~Ling, A.~Kurzawski, J.~Templeton, Reynolds averaged turbulence modelling
	using deep neural networks with embedded invariance, Journal of Fluid
	Mechanics 807 (2016) 155--166.
	
	\bibitem{duraisamy2019turbulence}
	K.~Duraisamy, G.~Iaccarino, H.~Xiao, Turbulence modeling in the age of data,
	Annual Review of Fluid Mechanics 51 (2019) 357--377.
	
	\bibitem{wang2017physics}
	J.-X. Wang, J.-L. Wu, H.~Xiao, Physics-informed machine learning approach for
	reconstructing reynolds stress modeling discrepancies based on dns data,
	Physical Review Fluids 2~(3) (2017) 034603.
	
	\bibitem{wang2019prediction}
	J.-X. Wang, J.~Huang, L.~Duan, H.~Xiao, Prediction of reynolds stresses in
	high-mach-number turbulent boundary layers using physics-informed machine
	learning, Theoretical and Computational Fluid Dynamics 33~(1) (2019) 1--19.
	
	\bibitem{zafar2020convolutional}
	M.~I. Zafar, H.~Xiao, M.~M. Choudhari, F.~Li, C.-L. Chang, P.~Paredes,
	B.~Venkatachari, Convolutional neural network for transition modeling based
	on linear stability theory, arXiv preprint arXiv:2005.02599.
	
	\bibitem{fukami2019synthetic}
	K.~Fukami, Y.~Nabae, K.~Kawai, K.~Fukagata, Synthetic turbulent inflow
	generator using machine learning, Physical Review Fluids 4~(6) (2019) 064603.
	
	\bibitem{kim2020deep}
	J.~Kim, C.~Lee, Deep unsupervised learning of turbulence for inflow generation
	at various reynolds numbers, Journal of Computational Physics 406 (2020)
	109216.
	
	\bibitem{sun2020surrogate}
	L.~Sun, H.~Gao, S.~Pan, J.-X. Wang, Surrogate modeling for fluid flows based on
	physics-constrained deep learning without simulation data, Computer Methods
	in Applied Mechanics and Engineering 361 (2020) 112732.
	
	\bibitem{maulik2020probabilistic}
	R.~Maulik, K.~Fukami, N.~Ramachandra, K.~Fukagata, K.~Taira, Probabilistic
	neural networks for fluid flow surrogate modeling and data recovery, Physical
	Review Fluids 5~(10) (2020) 104401.
	
	\bibitem{gao2019non}
	H.~Gao, J.-X. Wang, M.~J. Zahr, Non-intrusive model reduction of large-scale,
	nonlinear dynamical systems using deep learning, arXiv preprint
	arXiv:1911.03808.
	
	\bibitem{ledig2017photo}
	C.~Ledig, L.~Theis, F.~Husz{\'a}r, J.~Caballero, A.~Cunningham, A.~Acosta,
	A.~Aitken, A.~Tejani, J.~Totz, Z.~Wang, et~al., Photo-realistic single image
	super-resolution using a generative adversarial network, in: Proceedings of
	the IEEE conference on computer vision and pattern recognition, 2017, pp.
	4681--4690.
	
	\bibitem{fukami2020machine}
	K.~Fukami, K.~Fukagata, K.~Taira, Machine learning based spatio-temporal super
	resolution reconstruction of turbulent flows, arXiv preprint
	arXiv:2004.11566.
	
	\bibitem{fukami2019super}
	K.~Fukami, K.~Fukagata, K.~Taira, Super-resolution analysis with machine
	learning for low-resolution flow data, in: 11th International Symposium on
	Turbulence and Shear Flow Phenomena, TSFP 2019, 2019.
	
	\bibitem{fukami2018super}
	K.~Fukami, K.~Fukagata, K.~Taira, Super-resolution reconstruction of turbulent
	flows with machine learning, arXiv preprint arXiv:1811.11328.
	
	\bibitem{liu2020multiresolution}
	Y.~Liu, C.~Ponce, S.~L. Brunton, J.~N. Kutz, Multiresolution convolutional
	autoencoders, arXiv preprint arXiv:2004.04946.
	
	\bibitem{deng2019super}
	Z.~Deng, C.~He, Y.~Liu, K.~C. Kim, Super-resolution reconstruction of turbulent
	velocity fields using a generative adversarial network-based artificial
	intelligence framework, Physics of Fluids 31~(12) (2019) 125111.
	
	\bibitem{bode2019using}
	M.~Bode, M.~Gauding, Z.~Lian, D.~Denker, M.~Davidovic, K.~Kleinheinz,
	J.~Jitsev, H.~Pitsch, Using physics-informed super-resolution generative
	adversarial networks for subgrid modeling in turbulent reactive flows, arXiv
	preprint arXiv:1911.11380.
	
	\bibitem{bai2019dynamic}
	K.~Bai, W.~Li, M.~Desbrun, X.~Liu, Dynamic upsampling of smoke through
	dictionary-based learning, arXiv preprint arXiv:1910.09166.
	
	\bibitem{gonzalez2018deep}
	F.~J. Gonzalez, M.~Balajewicz, Deep convolutional recurrent autoencoders for
	learning low-dimensional feature dynamics of fluid systems, arXiv preprint
	arXiv:1808.01346.
	
	\bibitem{xie2018tempogan}
	Y.~Xie, E.~Franz, M.~Chu, N.~Thuerey, tempogan: A temporally coherent,
	volumetric gan for super-resolution fluid flow, ACM Transactions on Graphics
	(TOG) 37~(4) (2018) 1--15.
	
	\bibitem{liu2020deep}
	B.~Liu, J.~Tang, H.~Huang, X.-Y. Lu, Deep learning methods for super-resolution
	reconstruction of turbulent flows, Physics of Fluids 32~(2) (2020) 025105.
	
	\bibitem{werhahn2019multi}
	M.~Werhahn, Y.~Xie, M.~Chu, N.~Thuerey, A multi-pass gan for fluid flow
	super-resolution, Proceedings of the ACM on Computer Graphics and Interactive
	Techniques 2~(2) (2019) 1--21.
	
	\bibitem{guo2020ssr}
	L.~Guo, S.~Ye, J.~Han, H.~Zheng, H.~Gao, D.~Z. Chen, J.-X. Wang, C.~Wang,
	Ssr-vfd: Spatial super-resolution for vector field data analysis and
	visualization, in: 2020 IEEE Pacific Visualization Symposium (PacificVis),
	IEEE, 2020, pp. 71--80.
	
	\bibitem{erichson2019shallow}
	N.~B. Erichson, L.~Mathelin, Z.~Yao, S.~L. Brunton, M.~W. Mahoney, J.~N. Kutz,
	Shallow learning for fluid flow reconstruction with limited sensors and
	limited data, arXiv preprint arXiv:1902.07358.
	
	\bibitem{ferdian20204dflownet}
	E.~Ferdian, A.~Suinesiaputra, D.~J. Dubowitz, D.~Zhao, A.~Wang, B.~Cowan, A.~A.
	Young, {4DFlowNet}: Super-resolution 4d flow {MRI} using deep learning and
	computational fluid dynamics, Frontiers in Physics 8 (2020) 138.
	
	\bibitem{raissi2019physics}
	M.~Raissi, P.~Perdikaris, G.~Karniadakis, Physics-informed neural networks: A
	deep learning framework for solving forward and inverse problems involving
	nonlinear partial differential equations, Journal of Computational Physics
	378 (2019) 686--707.
	
	\bibitem{dwivedi420distributed}
	V.~Dwivedi, N.~Parashar, B.~Srinivasan, Distributed learning machines for
	solving forward and inverse problems in partial differential equations,
	Neurocomputing 420  299--316.
	
	\bibitem{zhu2019physics}
	Y.~Zhu, N.~Zabaras, P.-S. Koutsourelakis, P.~Perdikaris, Physics-constrained
	deep learning for high-dimensional surrogate modeling and uncertainty
	quantification without labeled data, Journal of Computational Physics 394
	(2019) 56--81.
	
	\bibitem{geneva_modeling_2020}
	N.~Geneva, N.~Zabaras,
	\href{https://linkinghub.elsevier.com/retrieve/pii/S0021999119307612}{Modeling
		the dynamics of {PDE} systems with physics-constrained deep auto-regressive
		networks}, Journal of Computational Physics 403 (2020) 109056.
	\newblock \href {http://dx.doi.org/10.1016/j.jcp.2019.109056}
	{\path{doi:10.1016/j.jcp.2019.109056}}.
	\newline\urlprefix\url{https://linkinghub.elsevier.com/retrieve/pii/S0021999119307612}
	
	\bibitem{zhang2020physics}
	R.~Zhang, Y.~Liu, H.~Sun, Physics-informed multi-lstm networks for metamodeling
	of nonlinear structures, arXiv preprint arXiv:2002.10253.
	
	\bibitem{long2017pde}
	Z.~Long, Y.~Lu, X.~Ma, B.~Dong, {PDE-net}: Learning {PDE}s from data, arXiv
	preprint arXiv:1710.09668.
	
	\bibitem{long2019pde}
	Z.~Long, Y.~Lu, B.~Dong, {PDE-Net} 2.0: Learning {PDE}s from data with a
	numeric-symbolic hybrid deep network, Journal of Computational Physics 399
	(2019) 108925.
	
	\bibitem{singh2020time}
	G.~Singh, S.~Gupta, M.~Lease, C.~N. Dawson, {TIME}: A transparent,
	interpretable, model-adaptive and explainable neural network for dynamic
	physical processes, arXiv preprint arXiv:2003.02426.
	
	\bibitem{chen2020deep}
	Z.~Chen, Y.~Liu, H.~Sun, Deep learning of physical laws from scarce data, arXiv
	preprint arXiv:2005.03448.
	
	\bibitem{jiang2020meshfreeflownet}
	C.~M. Jiang, S.~Esmaeilzadeh, K.~Azizzadenesheli, K.~Kashinath, M.~Mustafa,
	H.~A. Tchelepi, P.~Marcus, A.~Anandkumar, et~al., Meshfreeflownet: A
	physics-constrained deep continuous space-time super-resolution framework,
	arXiv preprint arXiv:2005.01463.
	
	\bibitem{mohan2020embedding}
	A.~T. Mohan, N.~Lubbers, D.~Livescu, M.~Chertkov, Embedding hard physical
	constraints in neural network coarse-graining of 3d turbulence, arXiv
	preprint arXiv:2002.00021.
	
	\bibitem{thuereyphysics}
	N.~Thuerey, Y.~Xie, M.~Chu, S.~Wiewel, L.~Prantl, Physics-based deep learning
	for fluid flow.
	
	\bibitem{subramaniam2020turbulence}
	A.~Subramaniam, M.~L. Wong, R.~D. Borker, S.~Nimmagadda, S.~K. Lele, Turbulence
	enrichment using physics-informed generative adversarial networks, arXiv
	(2020) arXiv--2003.
	
	\bibitem{sun2020physics}
	L.~Sun, J.-X. Wang, Physics-constrained bayesian neural network for fluid flow
	reconstruction with sparse and noisy data, arXiv preprint arXiv:2001.05542.
	
	\bibitem{SUN2019112732}
	L.~Sun, H.~Gao, S.~Pan, J.-X. Wang, Surrogate modeling for fluid flows based on
	physics-constrained deep learning without simulation data, Computer Methods
	in Applied Mechanics and Engineering\href
	{http://dx.doi.org/https://doi.org/10.1016/j.cma.2019.112732}
	{\path{doi:https://doi.org/10.1016/j.cma.2019.112732}}.
	
	\bibitem{gao2020phygeonet}
	H.~Gao, L.~Sun, J.-X. Wang, Phygeonet: Physics-informed geometry-adaptive
	convolutional neural networks for solving parametric pdes on irregular
	domain, arXiv preprint arXiv:2004.13145.
	
	\bibitem{shi2016real}
	W.~Shi, J.~Caballero, F.~Husz{\'a}r, J.~Totz, A.~P. Aitken, R.~Bishop,
	D.~Rueckert, Z.~Wang, Real-time single image and video super-resolution using
	an efficient sub-pixel convolutional neural network, in: Proceedings of the
	IEEE conference on computer vision and pattern recognition, 2016, pp.
	1874--1883.
	
	\bibitem{kingma2014adam}
	D.~P. Kingma, J.~Ba, Adam: A method for stochastic optimization, arXiv preprint
	arXiv:1412.6980.
	
	\bibitem{nair2010rectified}
	V.~Nair, G.~E. Hinton, Rectified linear units improve restricted boltzmann
	machines, in: ICML, 2010.
	
	\bibitem{johnson2010improved}
	K.~M. Johnson, M.~Markl, Improved snr in phase contrast velocimetry with
	five-point balanced flow encoding, Magnetic Resonance in Medicine: An
	Official Journal of the International Society for Magnetic Resonance in
	Medicine 63~(2) (2010) 349--355.
	
	\bibitem{jasak2007openfoam}
	H.~Jasak, A.~Jemcov, Z.~Tukovic, et~al., Openfoam: A c++ library for complex
	physics simulations, in: International workshop on coupled methods in
	numerical dynamics, Vol. 1000, IUC Dubrovnik Croatia, 2007, pp. 1--20.
	
	\bibitem{pletcher2012computational}
	R.~H. Pletcher, J.~C. Tannehill, D.~Anderson, Computational fluid mechanics and
	heat transfer, CRC press, 2012.
	
	\bibitem{rhie1983numerical}
	C.~Rhie, W.~L. Chow, Numerical study of the turbulent flow past an airfoil with
	trailing edge separation, AIAA journal 21~(11) (1983) 1525--1532.
	
	\bibitem{paszke2017automatic}
	A.~Paszke, S.~Gross, S.~Chintala, G.~Chanan, E.~Yang, Z.~DeVito, Z.~Lin,
	A.~Desmaison, L.~Antiga, A.~Lerer, Automatic differentiation in pytorch.
	
\end{thebibliography}

\end{document}